\documentclass[12pt,preprint]{aastex}
\usepackage{natbib}
\usepackage{url}
\usepackage{lscape}
\begin{document}
\title{Atmospheric Characterization of 5 Hot Jupiters with Wide Field Camera 3 on the Hubble Space Telescope}
\author{Sukrit Ranjan\altaffilmark{1,6}, David Charbonneau\altaffilmark{1}, Jean-Michel D\'{e}sert\altaffilmark{2}, Nikku Madhusudhan\altaffilmark{3}, Drake Deming\altaffilmark{4}, Ashlee Wilkins\altaffilmark{4}, and Avi M. Mandell\altaffilmark{5}}

\altaffiltext{1}{Harvard-Smithsonian Center for Astrophysics, Cambridge, MA 02138, USA}
\altaffiltext{2}{Department of Astrophysical and Planetary Sciences, University of Colorado, Boulder, CO 80309, USA}
\altaffiltext{3}{Institute of Astronomy, University of Cambridge, Cambridge CB3 0HA, UK}
\altaffiltext{4}{Department of Astronomy, University of Maryland, College Park, MD 20742, USA}
\altaffiltext{5}{NASA's Goddard Space Flight Center, Greenbelt, MD 20771, USA}
\altaffiltext{6}{sranjan@cfa.harvard.edu}

\date{\today}

\begin{abstract}
We probe the structure and composition of the atmospheres of 5 hot Jupiter exoplanets using the Hubble Space Telescope Wide Field Camera 3 (WFC3) instrument. We use the G141 grism (1.1-1.7 $\mu$m) to study TrES-2b, TrES-4b, and CoRoT-1b in transit, TrES-3b in secondary eclipse, and WASP-4b in both. This wavelength region includes a predicted absorption feature from water at 1.4 $\mu$m, which we expect to be nondegenerate with the other molecules that are likely to be abundant for hydrocarbon-poor (e.g. solar composition) hot Jupiter atmospheres. We divide our wavelength regions into 10 bins. For each bin we produce a spectrophotometric light curve spanning the time of transit and/or eclipse.  We correct these light curves for instrumental systematics without reference to an instrument model.  For our transmission spectra, our mean $1-\sigma$ precision per bin corresponds to variations of 2.1, 2.8,  and 3.0 atmospheric scale heights for TrES-2b, TrES-4b, and CoRoT-1b, respectively. We find featureless spectra for  these three planets. We are unable to extract a robust transmission spectrum for WASP-4b. For our dayside emission spectra, our mean $1-\sigma$ precision per bin corresponds to a planet-to-star flux ratio of $1.5\times10^{-4}$ and $2.1\times10^{-4}$ for WASP-4b and TrES-3b, respectively. We combine these estimates with previous broadband measurements and conclude that for both planets isothermal atmospheres are disfavored. We find no signs of features due to water. We confirm that WFC3 is suitable for studies of transiting exoplanets, but in staring mode multi-visit campaigns are necessary to place strong constraints on water abundance. 
\end{abstract}


\maketitle

\maketitle

\section{Introduction}
Transiting exoplanets offer unique opportunities for characterization of their atmospheres. Studying the wavelength-dependent depth of planetary transits and eclipses permits constraints on planetary composition and atmospheric structure. These techniques have been applied broadly in studies of hot Jupiters. Transmission spectroscopy was first used to detect atomic sodium \citep{Charbonneau2002a}, and later used to report the detection of molecules such as water \citep{Barman2007} and planetary hazes \citep{Pont2008}. For a detailed review, see \citet{Seager2010}. Broad-band thermal emission studies with Spitzer have enabled the study of thermal inversions and constrained the atmospheric redistribution of energy (see, for example, \citealt{Knutson2008} and \citealt{Desert2011a}). However, there persist challenges in the interpretation of similar data. The broad photometric bands used in many of these studies span multiple absorbers, leading to degeneracies in the interpretation. \citet{Madhusudhan2010a} show that there further exist degeneracies between composition and thermal structure: for example, they showed that the broad-band emission spectrophotometry of TrES-2b \citep{O'Donovan2006} and TrES-4b \citep{Mandushev2007} can be matched with or without thermal inversions, depending on the assumed composition.  Further spectroscopic observations in other wavelength regimes are required to break these degeneracies.  \citet{Madhusudhan2010a} highlight the potential of spectroscopy in the NIR for such studies owing to the abundant molecular absorption features in this wavelength regime.

Space-based NIR transit studies were previously performed with the Near-Infrared Camera and Multi-Object Spectrometer (NICMOS) instrument on the Hubble Space Telescope (HST). Using this instrument, \citet{Swain2008} reported the detection of H$_2$O, and CH$_4$ in the atmosphere of HD189733b, and \citet{Tinetti2010} reported the detection of CO$_2$, H$_2$O, and CH$_4$ in XO-1b . The NICMOS data contained strong systematics and it was necessary to decorrelate the data using a linear function of optical state vectors. \cite{Gibson2011a} explored alternate means for decorrelating the data by reanalyzing past NICMOS datasets. They experimented with using different out-of-transit orbits to decorrelate the data, using a quadratic instead of linear function for decorrelation, and altering the set of parameters used for decorrelation. They found the resulting shape of the extracted spectrum to depend on the decorrelation technique, and they could not confirm the molecules reported by previous studies. \citet{Crouzet2012} analyzed NICMOS archival data of XO-1b, and concluded the uncertainty in correcting for NICMOS instrument systematics was comparable to expected variations due to atmospheric absorption. In a follow-up study, \citet{Gibson2012} used Gaussian processes to analyze the HD189733b NICMOS data and extracted a spectrum that was consistent with \citet{Swain2008}. However, the uncertainties on this spectrum were much higher, and \citet{Gibson2012} found there was no strong evidence for the molecular detections reported by \citet{Swain2008}. However, it is important to note that this perspective is still debated by the original authors \citep{Swain2011}.  \citet{Waldmann2013} use an independent component analysis (ICA) on the NICMOS data to derive a spectrum with uncertainties intermediate to \citet{Swain2008} and \citet{Gibson2012}. \citet{Swain2014} conduct a uniform Bayesian model comparison using multiple retrieval algorithms of these spectra, and find that \citet{Swain2008} and \citet{Waldmann2013} are consistent with molecular detections, but \citet{Gibson2012} is not. Follow-up observations are required to validate or refute the reported detections with NICMOS. 

The Wide-Field Camera 3 (WFC3) \citep{Dressel2012} is the only space-based NIR spectrometer currently in operation suitable for transiting exoplanet observations. The G141 grism on HST is sensitive to 1.1-1.7 $\mu$m. This wavelength range spans a water-absorption feature at 1.4 microns. For solar composition (oxygen rich) hot Jupiters, this feature is not degenerate with any other major predicted atmospheric absorber. The grism also spans an atmospheric window at 1.6 $\mu$m, where no absorption is predicted. Hence measurements here probe the photospheric emission from the planet and may constrain the planetary energy budget. The wavelength coverage of the WFC3 G141 grism largely overlaps with the 1.2-1.8 $\mu$m range of the NICMOS G141 grism, meaning studies of hot Jupiter atmospheres with WFC3 can test claims based on NICMOS data. For example, \citet{Swain2014} point out that the water abundances predicted for HD189733b by the results of \citet{Waldmann2013} and \citet{Swain2008} should produce a 300-400 ppm signature in the WFC3 IR bandpass. Compared to NICMOS, WFC3 has higher throughput\footnote{See \citet{Dressel2012}, specifically \url{http://www.stsci.edu/hst/wfc3/documents/handbooks/currentIHB/c03_optimum_instr4.html}}, and ground-test studies indicate WFC3 is characterized by more uniform intra-pixel sensitivity response than is NICMOS \citep{McCullough2008}. Observations by the higher-performance WFC3, coupled with ground-based studies (e.g. \citealt{Danielski2014}), offer a means to test reported detections with NICMOS.

In this paper, we present results from a study of 5 hot Jupiter atmospheres using WFC3. A comparative study with the same instrument minimizes the risk that instrumental effects or choices in the analysis are introducing systematic differences between spectra, and thus may permit us to begin to make statements about hot Jupiters as a class. Our program is part of a larger study of 16 hot Jupiters. The first results of the larger study were released by \citet{Deming2013} and \citet{Mandell2013}. \citet{Deming2013} presented observations of XO-1b and HD209458b in HST drift-scan mode,  whereby 'nodding' the telescope over the course of an exposure alleviates data gaps due to buffer dumps and improves the duty cycle.The observations of WASP-12b, WASP-17b, and WASP-19b presented by  \citet{Mandell2013}, by contrast, were obtained in the 'staring' imaging mode used by \citet{Berta2011}.The observations presented in this paper were similarly obtained in staring mode.

\begin{figure}[h]
\centering
\includegraphics[width=15 cm, angle=0]{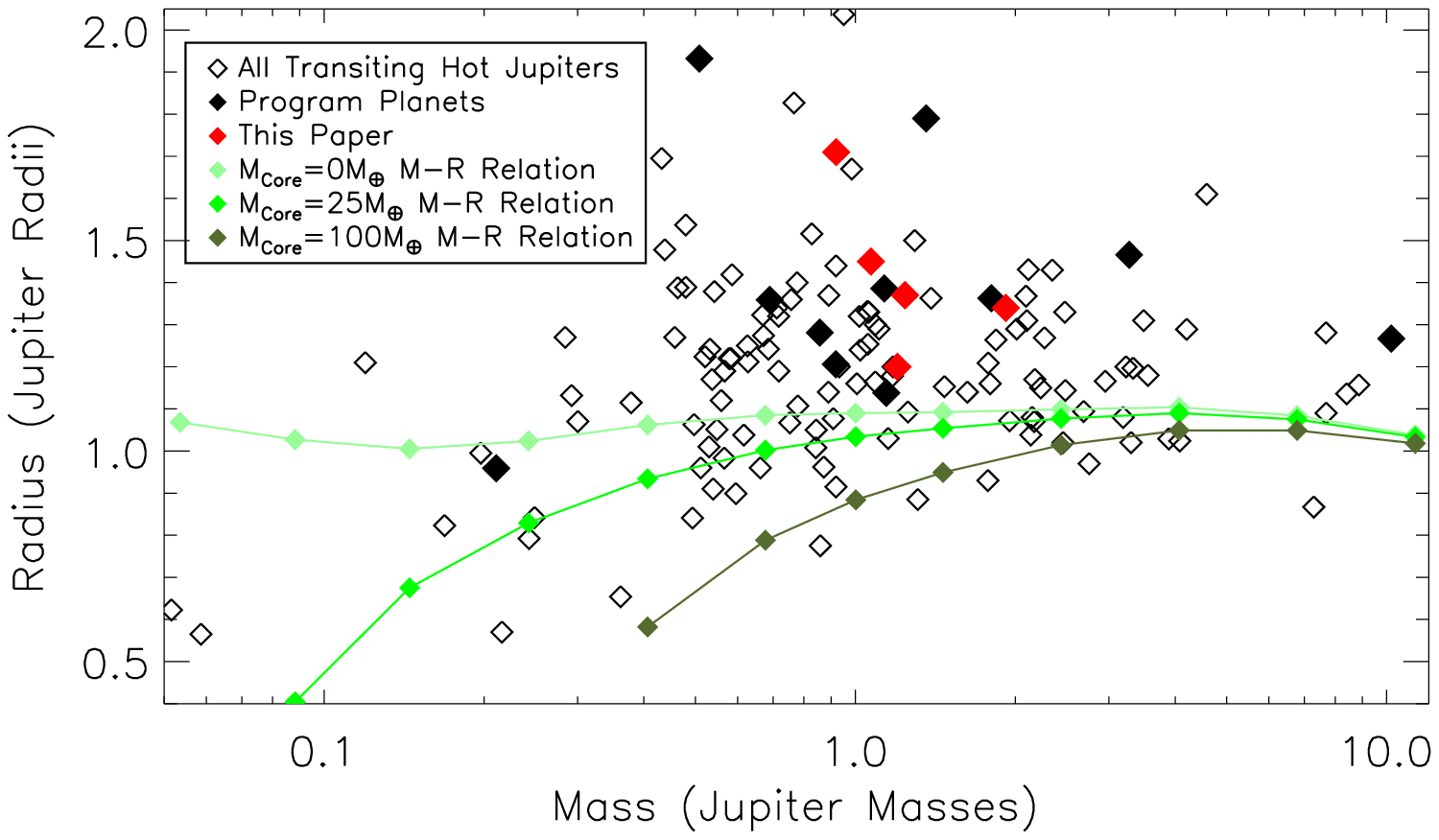}
\caption{Mass-radius diagram for the known transiting Jovian planets ($R>0.5R_{\rm{Jup}}$) with both mass and radius measurements. The planets in the WFC3 program are filled-in black diamonds. The planets analyzed in this paper are filled-in red diamonds. This figure was generated using planet parameters from the Exoplanet Data Explorer, \url{http://exoplanets.org/} \citep{Wright2011}, except for the parameters for the planets in this study, which are from the references in Table~\ref{tbl:targetparams}. We overplot theoretical mass-radius relations from \citet{Fortney2007} in green. The relations plotted are for 4.5-Gyr-old giant hydrogen-helium planets (helium mass fraction $Y=0.28$) orbiting a Solar-type star at 0.1 AU, for heavy element cores of mass of $0 M_{\Earth}$ (pale green), $25 M_{\Earth}$ (green), and $100 M_{\Earth}$ (olive). These relations are provided for a sense of scale only, and should not be taken to rule on whether the exoplanets presented here are bloated, since such a judgement requires detailed modeling involving parameters beyond the mass and radius.\label{fig:massradius}}
\end{figure}

This work presents results from TrES-2b, TrES-4b, and CoRoT-1b \citep{Barge2008} in transit, TrES-3b \citep{Odonovan2007} in secondary eclipse, and WASP-4b \citep{Wilson2008} in both. Table~\ref{tbl:targetparams} summarizes the physical parameters of these objects, and Figure~\ref{fig:massradius} summarizes these worlds in the context of the known transiting Jovian planets. These 5 planets were chosen for their accessibility (from a signal-to-noise perspective) to transit and eclipse studies, resulting from their short-period orbits, large radii, and high effective temperatures. Our sample includes planets with and without reported thermal inversions from previous measurements, as well as  "bloated" hot Jupiters, which are gas giants whose radii are larger than expected from conventional equilibrium models. The degree to which the measured radii of the 5 planets disagree with the expectations from structural models varies, from nearly in agreement (TrES-2b), to observed radii that are substantially larger than predicted (TrES-4b). Measurements of the planetary energy budget and composition via the thermal emission and transmission spectroscopy performed in this paper can inform the physical explanation for their variation in physical radii.

\begin{table}[h]
\begin{center}
\caption{Physical parameters of observed exoplanets and their hosts. \label{tbl:targetparams}}
\begin{tabular}{p{1.9 cm}p{1.1cm}p{1.1cm}p{1.1cm}p{0.9cm}p{0.9cm}p{0.7cm}p{0.7cm}p{1.0cm}p{4.5cm}}
\tableline\tableline
Planet & $M_P$ ($M_{\rm{Jup}}$)& $R_P$ ($R_{\rm{Jup}}$) & Period (Days) & $M_\star$ ($M_{\Sun}$)& Sp. Type & $T_{\rm{eq,P}}$\tablenotemark{a} (K) & $T_{\rm{eff,\star}}$ (K)& $H/R_P$\tablenotemark{b} &References \\
\tableline
TrES-2b  & 1.20 & 1.20 & 2.47 & 0.992 & G0V & 1700 &5850& 0.0036 & \citet{Kipping2011a}\\
TrES-3b   & 1.92 & 1.34 & 1.31 & 0.928 & K0V & 2000 &5650&0.0030& \citet{Sozzetti2009}\\
TrES-4b  & 0.92 & 1.71 & 3.55 & 1.388 & F8V & 2100 &6200&0.0083& \citet{Chan2011}\\
WASP-4b & 1.24 & 1.37 & 1.34 & 0.925 & G7V & 2000 &5500& 0.0047& \citet{Gillon2009}, \citet{Winn2009}\\
CoRoT-1b & 1.07 & 1.45 & 1.51 & 1.01 & G0V & 2200 &5950&0.0063&\citet{Gillon2009a}, \citet{Barge2008}\\
\tableline
\end{tabular}
\tablenotetext{a}{Planetary equilibrium temperature estimate. Assumes blackbody stellar emission at $T_{\rm{eff,\star}}$ and zero albedo atmospheres with no redistribution of energy, and are thus meant only to be comparative.}
\tablenotetext{b}{Atmospheric scale height, computed by $H=kT/\mu g$, where  $T=T_{\rm{eq,P}}$, $g=GM_P/R_P^2$, and the mean molecular mass $\mu=2.20m_p$, where $m_p$ is the proton mass.}
\end{center}
\end{table}

The paper is organized as follows. In Section~\ref{sec:obs}, we describe the collection of our data. In Section~\ref{sec:extract}, we describe the reduction of our data and the removal of instrumental systematics. In Section~\ref{sec:fitting}, we describe how we fit transit and eclipse models to the data, and in Section~\ref{sec:validation} we describe tests we performed to assess the robustness of our analysis. We interpret the resulting transmission spectra in Section~\ref{sec:prelimres}, and summarize our conclusions in Section~\ref{sec:conc}.

\section{Observations\label{sec:obs}}
\begin{figure}[h]
\centering
\includegraphics[width=18 cm, angle=0]{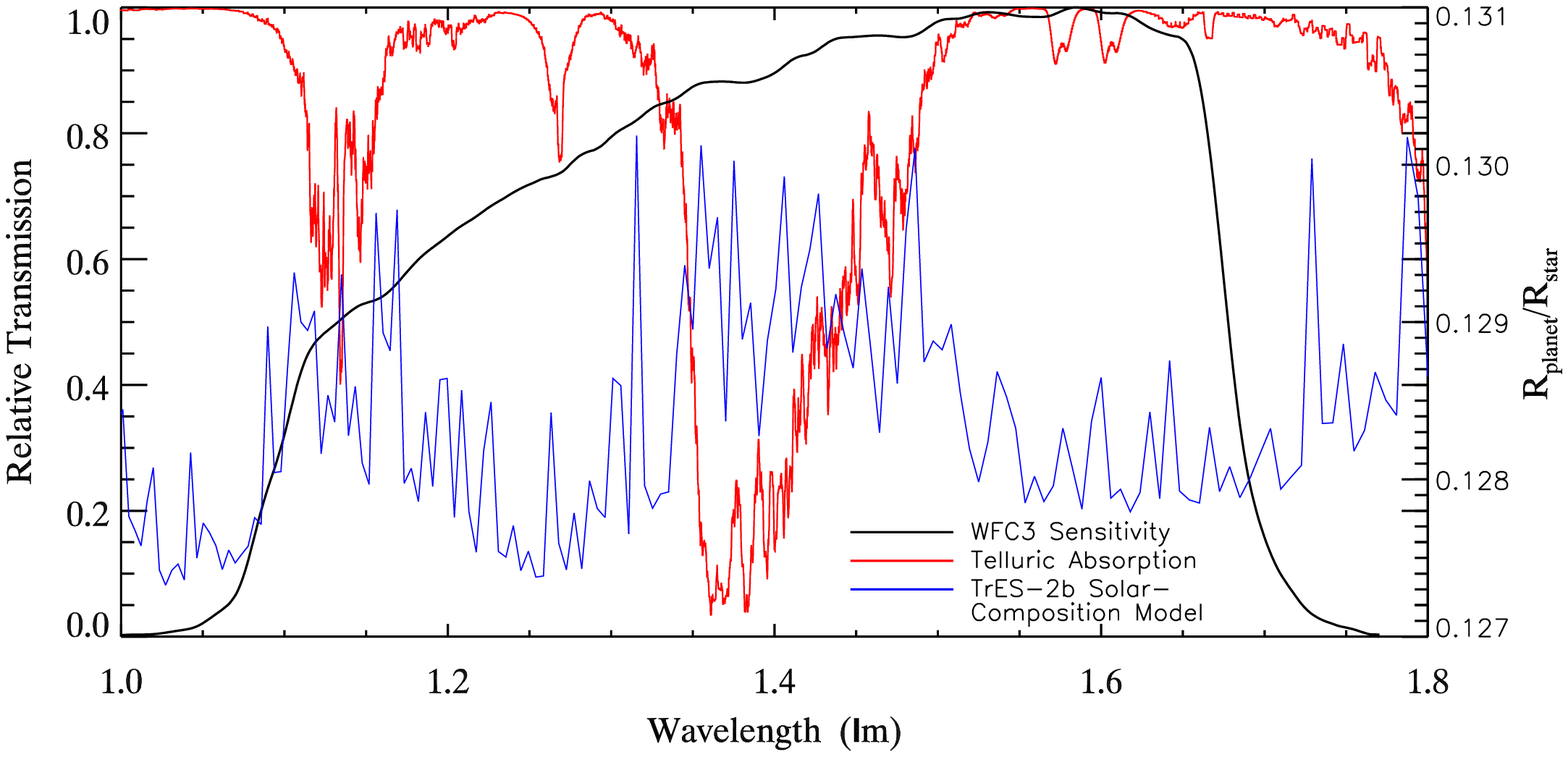}
\caption{WFC3 relative transmission with the G141 grism (black), coplotted with telluric atmospheric transmission (red). Also plotted is a nominal model solar-composition transmission spectrum for TrES-2b showing variations in apparent planet radius with wavelength. Note the predicted H$_2$O absorption feature at 1.4 microns. Telluric absorption at this feature mean that it is best studied from space. \label{fig:wfc3g141response}}
\end{figure}

The IR channel of WFC3 pairs a 1024 $\times$ 1024 HgCdTe detector with a selection array of 15 filters and two grisms. We used the G141 grism, spanning 1.1-1.7 $\mu$m at $R=\lambda/\Delta\lambda\approx130$\footnote{See \url{http://www.stsci.edu/hst/wfc3/analysis/grism_obs/wfc3-grism-resources.html}}. Table~\ref{tbl:obs} summarizes our observations, and Figure~\ref{fig:wfc3g141response} shows the instrument throughput\footnote{From \url{http://www.stsci.edu/hst/wfc3/analysis/grism_obs/calibrations/wfc3_g141.html}}  coplotted with telluric transmission\footnote{From \url{ftp://ftp.noao.edu/catalogs/atmospheric_transmission/}}. The study of the putative 1.4 $\mu$m water band is best conducted from space as telluric absorption would interfere with ground-based observation. 

We observed each event (transit/eclipse) with an HST visit of 4 or 5 orbits. At the time of our campaign, HST orbits were 96 minutes long, with 45-minute data gaps during each orbit due to terrestrial occultation. Pointing the telescope at a new target changes the insolation of the spacecraft, and it takes some time for the spacecraft and its optics to settle back into an equilibrium state. Hence, the first orbit of a visit usually displays unique systematics. Consequently, visits were constructed to have a  first orbit that could be discarded, followed by an orbit before the event, 1-2 orbits spanning the event, and in 3 of the 5 cases another orbit after the event. Instrument overheads are dominated by buffer-read out time. It takes 5.8 minutes to read out the full 1024 $\times$ 1024 pixel image, and WFC3 can only hold the equivalent of 2 full-frame exposures in memory \citep{Berta2011}.  In order to optimize photon collection efficiency, only a subarray of the detector corresponding to the target spectrum was read out, reducing the number of buffer reads per orbit. Additionally, we chose exposure times close to nonlinearity (saturation). Even at saturation an unsaturated signal can be recovered, because WFC3 IR exposures are "sampled up the ramp", with multiple nondestructive reads collected per exposure. Saturated reads are flagged and rejected by the WFC3 pipeline, and not used in estimating flux rates \citep{Dressel2012, Rajan2011}. 

 Table~\ref{tbl:obs} gives the exposure times, readout mode and subarray size selected for each object. The wavelength solution outlined in \citet{Kuntschner2009} depends on the displacement between the direct image and the first-order spectrum, so for wavelength calibration a direct image was collected during each visit using the F139M filter. To avoid systematics from the detector flat fields, which have a precision of only $\approx0.5\%$ \citep{Pirzkal2011}, the pointing was not dithered. We observed drifts at the 0.03-0.06 pixel levels in our datasets. 
 
\begin{landscape}
\begin{table}[h]
\small
\begin{center}
\caption{Summary of observations used in this paper\label{tbl:obs}}
\begin{tabular}{cccccccccc}
\tableline\tableline
Event & $\#$Orbits & $V$ & $J$ &$H$ & Readout Mode & Subarray Size & NSAMP & Exposure Time & $N_{\rm{spectra}}$\\ 
 & & (mag) & (mag) & (mag) &  & (pixels) &  &  (s) & \\
\tableline
TrES-2 Transit & 4 &11.4 &10.2&9.9& RAPID &$512\times512$& 16 &12.80 &105\\ 
TrES-3 Eclipse & 4  &12.4 &11.0&10.7& SPARS10 & $128\times128$& 7 & 36.02& 219\\
TrES-4 Transit & 5 & 11.6 &10.6&10.4& RAPID &$512\times512$& 16 &12.80& 140\\
WASP-4 Transit & 5 & 12.5 &11.2 &10.8& SPARS10 &$128\times128$ & 7 & 36.02& 268\\
WASP-4 Eclipse & 5  & 12.5 &11.2 &10.8 & SPARS10 &$128\times128$ & 7 & 36.02&268\\
CoRoT-1 Transit & 4  & 13.6 &12.5&12.2& SPARS10 &$128\times128$ & 16 &100.65&98\\
\tableline
\end{tabular}
\end{center}
\end{table}
\end{landscape}

\section{Data Reduction\label{sec:extract}}
\subsection{Spectral Extraction}
We developed a data reduction pipeline to extract spectrophotometric light curves for each visit. We begin with the \emph{flt} image files output from the \emph{calwf3} pipeline. These images have undergone bias and dark current corrections, cosmic ray rejection, gain calibration, and been corrected for photometric nonlinearity, to compute the count rate per pixel. Saturated reads are flagged on a pixel-by-pixel basis in earlier files and not used to compute the count rates\footnote{For details, see \citealt{Dressel2012}, section 5.7.5 and \citealt{Rajan2011}, section 3.4.3}. We converted the Modified Julian Date $MJD=JD-2400000.5$ timestamp on each image to Barycentric Julian Date in the Barycentric Dynamical Time ($BJD_{TDB}$) to eliminate the effect of light travel time due to the Earth's orbit using the $utc2bjd.pro$ IDL code\footnote{\url{http://astroutils.astronomy.ohio-state.edu/time/}} described in \citet{Eastman2010a}. 

The spectra are closely aligned with the detector edge: The orientation angles of the spectra vary from $0.49^\circ \pm 0.02^\circ$ for the WASP-4b transit to $0.558^\circ \pm 0.009^\circ$ for TrES-3b. This alignment allows us to pursue a column-by-column extraction procedure. We use a box extraction centered on the spectral trace to extract a 1-dimensional spectrum. We estimated the trace center by choosing the detector row with maximum flux over the course of the visit. The height of the extraction box was chosen as follows: for each trial height, we extracted the light curve, corrected for systematics and fit a transit/eclipse curve (see Sections~\ref{sec:syscorr} and \ref{sec:fitting}). Taking steps of integer half-height, we then selected the height value that yielded the smallest formal error on $R_P/R_\star$.  Our extraction apertures varied but were typically 11 pixels in height. We experimented with another extraction method involving fitting individual Gaussians to the flux in each column and summing the flux spanned in a given extent of the Gaussian, for which several extents were tested. This method allowed the width of the spectrum to vary with column and image. For this method, we allowed for fractional pixels; for example, if we needed to integrate from pixels 1-12.4, we would sum pixels 1-12 and $0.4\times$ pixel 13. We found that the typical width of the Gaussian $\sigma<1$ pixel, indicating a tight PSF. This method returned spectra that were consistent within $1-\sigma$ with those derived using the box extraction. We elected to use the simpler box extraction technique to extract the 1-dimensional spectrum to enable ease of reproduction of our results. An example of an extracted spectrum is plotted in Figure~\ref{fig:treswavecalcorrected}.

In the case of WASP-4, TrES-3, and CoRoT-1 the entire $128\times128$ pixel image underwent this procedure as there were no contaminant stars in the aperture. In the case of TrES-2  and TrES-4, which were imaged in $512\times512$ mode, contaminant stars were present. To exclude these stars, we cut out sub-images and ran our extraction routine on them. Our sub-images were chosen to be as large as possible while being free of contaminant stars. The sub-image dimensions (x $\times$ y) were: 161$\times$83 for TrES-2 and 146$\times$95 for TrES-4. These apertures were chosen to include the full wavelength range of the detected spectrum while excluding other stars on the detector. For our datasets, the first-order spectrum had a dispersion of 0.0046 $\mu$m/pixel in the x-direction, and spatial full-width at half-maximums varied from 1.9-2.2 pixels.

\subsection{Background Subtraction\label{sec:backsub}}

We estimate and subtract off the background from our time-series spectra by choosing a static area on the detector free of object flux in each individual 2-D image. This area matches the wavelength range of the spectrum. These background columns are integrated and scaled to match the spectral extraction aperture on a column-by-column basis.  The width of the background aperture is tuned for each dataset to minimize scatter in the residuals of the white-light curve.  We tested dynamically integrating into the wings of the PSF on a column-by-column basis to estimate the background, and derived consistent results with the "static box" background estimate. We also considered the alternative method of flat-fielding the data\footnote{The \emph{flt} data are not flat-fielded.} and using histogram fitting to estimate the wavelength-independent background. Flat-fielding is required to correct for the different wavelength-dependent quantum efficiency function of each pixel. This method has the advantage of using more pixels to estimate the background compared to the "background cutout" method. We elected to use column-by-column subtraction as it is more robust to spectral variations in the background and any changes in the instrument state not accounted for by the flat-fields. We find the background as a fraction of stellar flux varies from $0.3\%$ (WASP-4) to $5\%$ (TrES-2). All objects besides TrES-2 have background of $<1\%$ of stellar flux. The background is not sensitive to variations in the background extraction box.

\subsection{Building Wavelength-Dependent Time Series}
We extract a 1-D background-subtracted spectrum from each data image, and order them in time to form a spectrophotometric time series. We bin our time series in wavelength to enhance signal-to-noise per resolution element. We choose wavelength bins such that each channel has the same number of photons over the course of the observations ("equal-photon binning"). We experimented with varying levels of binning, and found mutually consistent planetary spectra for numbers of bins ranging from 5-20.  Table~\ref{tbl:binboundaries} gives the boundaries of the wavelength bins for the data corresponding to each event for the final 10-channel spectra.

\begin{table}[h]
\small
\caption{Wavelength bin boundaries (in $\mu$m)  for each event.\label{tbl:binboundaries}}
\begin{tabular}{p{1.2cm}p{2.0cm}p{2.0cm}p{2.0cm}p{2.2cm}p{2.0cm}p{2.0cm}}
\tableline\tableline
Bin $\#$ & TrES-2 Transit  &  TrES-3 Eclipse &  TrES-4 Transit &  WASP-4 Transit  &  WASP-4 Eclipse &  CoRoT-1 Transit \\
\tableline
Bin 0 &   1.044-1.149 &  1.136-1.189 &   1.091-1.154 &   1.104-1.159 &   1.104-1.159 &  1.118-1.170 \\
Bin 1 &   1.149-1.205 &  1.189-1.237 &   1.154-1.207&   1.159-1.209 &   1.159-1.210 &  1.170-1.218 \\
Bin 2 &   1.205-1.257 &  1.237-1.284 &   1.207-1.257 &   1.209-1.256 &   1.210-1.257 &  1.218-1.264 \\
Bin 3 &   1.257-1.309 &  1.284-1.330 &   1.257-1.308 &   1.256-1.302 &   1.257-1.303 &  1.264-1.311 \\
Bin 4 &   1.309-1.361 &  1.330-1.375 &   1.308-1.358 &   1.302-1.347 &   1.303-1.348 &  1.311-1.357 \\
Bin 5 &   1.361-1.415 &  1.375-1.423 &   1.358-1.413 &   1.347-1.394 &   1.348-1.395 &  1.357-1.405 \\ 
Bin 6 &   1.415-1.472 &  1.423-1.471 &   1.413-1.469 &   1.394-1.442 &   1.395-1.443 &  1.405-1.455 \\
Bin 7 &   1.472-1.535 &  1.471-1.523 &   1.469-1.529 &   1.442-1.493 &   1.443-1.494 &  1.455-1.507 \\
Bin 8 &   1.535-1.603 &  1.523-1.578 &   1.529-1.596 &   1.493-1.547 &   1.494-1.547 &  1.507-1.561 \\   
Bin 9 &   1.603-1.704 &  1.578-1.638 &   1.596-1.680 &   1.547-1.606 &   1.547-1.606 &  1.561-1.619 \\   
\tableline
\end{tabular}
\end{table}

\subsection{Manual Exclusion of Data\label{sec:manexdata}}
As an initial check of our data, we plotted the white-light curves to check by eye for evidence of abnormalities that might be evidence of contamination due to effects such as cosmic rays. 4 points in the TrES-3b dataset were evident by eye to be highly discrepant from the eclipse light curve. We performed a $divide-oot$ correction as described in Section~\ref{sec:syscorr}, averaging the framing out-of-transit orbits and dividing the resultant vector into the in-transit orbit. The same 4 points were measured to be $>9 \sigma$ discrepant from the median flux observation on a per orbit basis, where the standard deviation was measured after excluding these outliers. The images corresponding to these data points were excluded entirely from analysis and deleted from the lightcurve. We checked for evidence of strong cosmic rays in these 2D frames but did not observe any. We ruled out the possibility of contamination due to the South Atlantic Anomaly (SAA) as the spacecraft was not near the SAA in any of these observations. The influence of these phenomena should be minimal since the $flt$ files have been cosmic ray rejected. 

In our analysis, we were concerned about the vulnerability of the data to edge effects. The edges contain substantially less flux. Since the instrument sensitivity function is most sharply sloped at the edges, the edges are also most vulnerable to the motion of the spectrum on the detector (drift). To mitigate this risk, we removed the 10 red-most and blue-most columns of our data. Minimal flux was lost due to this maneuver since the edges register substantially fewer photons.

\subsection{Correction of Systematics\label{sec:syscorr}}
\begin{figure}[h]
\centering
\includegraphics[width=15 cm, angle=0]{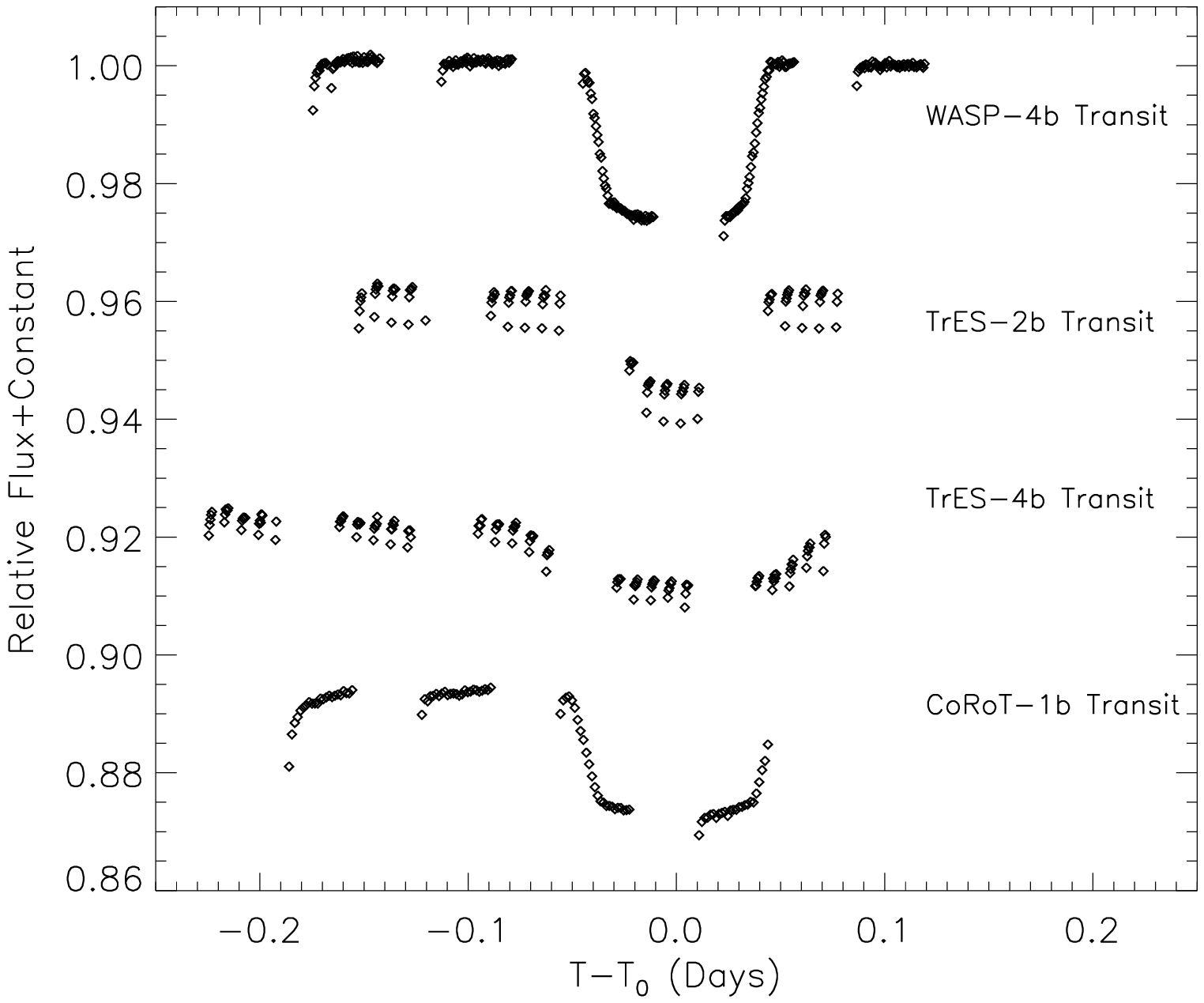}
\caption{Extracted light curves for the white-light transits observed. The light curves shown are, in order from top to bottom, those corresponding to WASP-4b, TrES-2b, TrES-4b, and CoRoT-1b, normalized to their maximal values and shifted vertically for display purposes. The time indicated is $T-T_0$, where $T_0$ is the epoch of transit listed in Table~\ref{tbl:systemparameters}. The flux indicated is the total flux integrated across the bandpass. The primary systematic is the ramp. Note the qualitatively different systematics observable in orbit 1, which is discarded in the analysis. \label{fig:uncorrtransit}}
\end{figure}

\begin{figure}[h]
\centering
\includegraphics[width=15 cm, angle=0]{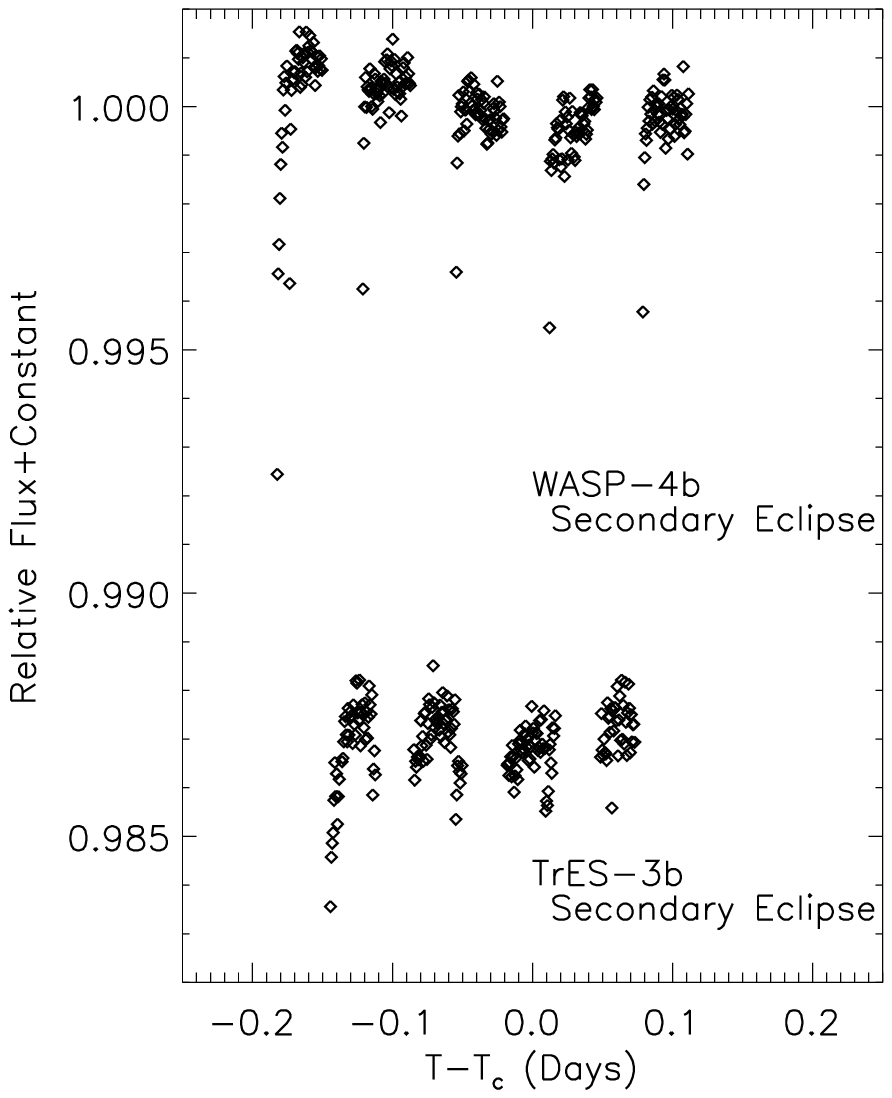}
\caption{Extracted light curves for the white-light secondary eclipses observed. The light curves shown are, in order from top to bottom, WASP-4b and TrES-3b, normalized to their maximal values and shifted vertically for display purposes. The time indicated is $T-T_c$, where $T_c$ is the epoch of eclipse listed in Table~\ref{tbl:systemparameters}. The flux indicated is the total flux integrated across the bandpass. The primary systematic is the ramp. Note the qualitatively different systematics observable in orbit 1, which is discarded in the analysis. \label{fig:uncorreclipse}}
\end{figure}

The dominant systematic observed is the sharply rising but quickly leveling ramp effect in flux described by \citet[see Figure~\ref{fig:uncorrtransit} and Figure~\ref{fig:uncorreclipse}]{Berta2011}\footnote{This is the same effect that \citet{Deming2013} call the "hook".}. That work suggests persistence as a possible cause of this systematic. We follow the \citet{Berta2011} $divide-oot$ method of correcting this systematic. We discard the first orbit's data (fluxes), which show unique systematics due to instrument resettling after repointing. From the remaining data, we interpolate the fluxes from leading and trailing out of event orbits and divide them into the in-event orbits as follows: consider a visit with five successive evenly-spaced orbits of data $u, w, x, y, z$. Let the vector of photometric time-series of fluxes corresponding to each successive orbit be labeled by $\vec{F_u}$, $\vec{F_w}$, $\vec{F_x}$, $\vec{F_y}$, and $\vec{F_z}$, with $x$ and $y$ being the in-event orbits. Then we would discard the fluxes from the first orbit $\vec{F_u}$ and correct the in-transit orbit fluxes to $\vec{F_{x}'}= \frac{\vec{F_x}}{0.75\vec{F_w}+0.25\vec{F_z}}$ and $\vec{F_y'}=\frac{\vec{F_y}}{0.25\vec{F_w}+0.75\vec{F_z}}$. In the case of a visit with four orbits $w$, $x$, $y$, and $z$, with only $y$ being in-transit, the fluxes corresponding to the first orbit $\vec{F_w}$ are discarded and the in-transit orbit is corrected to $\vec{F_y'}= \frac{\vec{F_y}}{0.5\vec{F_x}+0.5\vec{F_z}}$. This correction is done on a channel-by-channel wavelength basis.

In the case of WASP-4b and TrES-4b, the last orbit in the dataset that has one (TrES-4b) or two (WASP-4b) fewer points than the other non-discarded orbits. For this orbit, the $divide-oot$ procedure is modified. First, the fluxes from the orbit are padded by repeating the last value. $divide-oot$ is then carried out as above. Finally, the padding data points are removed. This procedure is reasonable since the ramp quickly levels off and the level of flux is essentially unchanging at the padded points. We checked the residuals corresponding to these padded points for anomalies and found none. 

Examining the light curves prior to detrending on a channel-by-channel basis for the different datasets, we observe wavelength-dependent systematics. In particular, we observe a visit-wide linear trend in flux. Both the sign and magnitude of the slope of this trend vary with wavelength channel; $divide-oot$ corrects this effect. This systematic illustrates the importance of correcting WFC3 light curves on a channel-by-channel basis (i.e. per wavelength bin).

We note that for two of our data sets, CoRoT-1b and TrES-4b, we do not have an HST orbit following the end of the event. We describe the treatment of these cases in Section~\ref{sec:singleoot}.

\subsection{Photometric error}

In estimating the errors on our measured fluxes, we consider the error that would result from photon noise. We measure the median photon level $p$ in a given spectrophotometric channel of a given orbit $y$ and assign the error on each point in the orbit to be $\sqrt{p}$. This approach underestimates the noise associated with the in-transit data: the $divide-oot$ method incorporates the scatter on the out-of-transit data into the in-transit data. To account for this, we scale up the errors on the in-transit orbit. Assuming random uncorrelated errors, we find that for a corrected orbit $\vec{F_y'}=\frac{\vec{F_y}}{a\vec{F_x}+(1-a)\vec{F_z}}$, $\sigma_{\vec{F_y'}}\approx\frac{\sigma_{\vec{F_y}}}{a\bar{F_x}+(1-a)\bar{F_z}}\sqrt{1+a^2+(1-a)^2}$. In the case of a=0.5, a simple average, this corresponds to an inflation by a factor $\sqrt{3/2}$. Comparing this noise estimate to the scatter of the residuals lets us estimate by how much the errors are affected by systematics (see Table~\ref{tbl:ppm} and Section~\ref{sec:fitting}).

\subsection{Outlier Removal\label{sec:sigclip}}
We implement sigma-clipping for outliers in each channel. We do an initial fit to OOT-corrected in-transit data and divide this initial fit into the data on a channel-by-channel basis. We compute the standard deviation $\sigma$ of the resultant vector. We flag points that are more than $3-\sigma$ from the model. Clipped points are replaced by the mean of their nearest neighbors. The datasets are well-behaved with few outliers: in bandpass-integrated white light 2 points are affected in the TrES-4b transit dataset. Depending on the dataset, we correct between $0-1.7\%$ of the data. We note that a second pass of the sigma clipping algorithm enhances the precision of the TrES-4b dataset in some channels. However, the second pass does not affect our spectrum or conclusions in any significant way. We elect to maintain a single pass of our clipping algorithm. 

\subsection{Wavelength Calibration}

We follow the method of \citet{Kuntschner2009} in determining our wavelength solution. The spectral solution is well-fit by the linear expression $\lambda=(\frac{dl}{dp})_0+(\frac{dl}{dp})_1*l$, where $l$ is the displacement between the direct image and the pixel of interest: $l=\sqrt{\Delta x^2+\Delta y^2}$. The values of  $(\frac{dl}{dp})_0$ and $(\frac{dl}{dp})_1$ are dependent on the location of the direct image $(x_{center}, y_{center})$. Their values are determined by an expansion around $(x_{center}, y_{center})$ using coefficients provided in \citet{Kuntschner2009}. We determine the location of the star in the F139M filter direct image via DS9.  Since the trace is horizontally aligned, $\Delta y \approx 0$ and we can write $\lambda=(\frac{dl}{dp})_0+(\frac{dl}{dp})_1*\Delta x$ and determine the wavelength fit on a column-by-column basis. 

\begin{figure}[h]
\centering
\includegraphics[width=15 cm, angle=0]{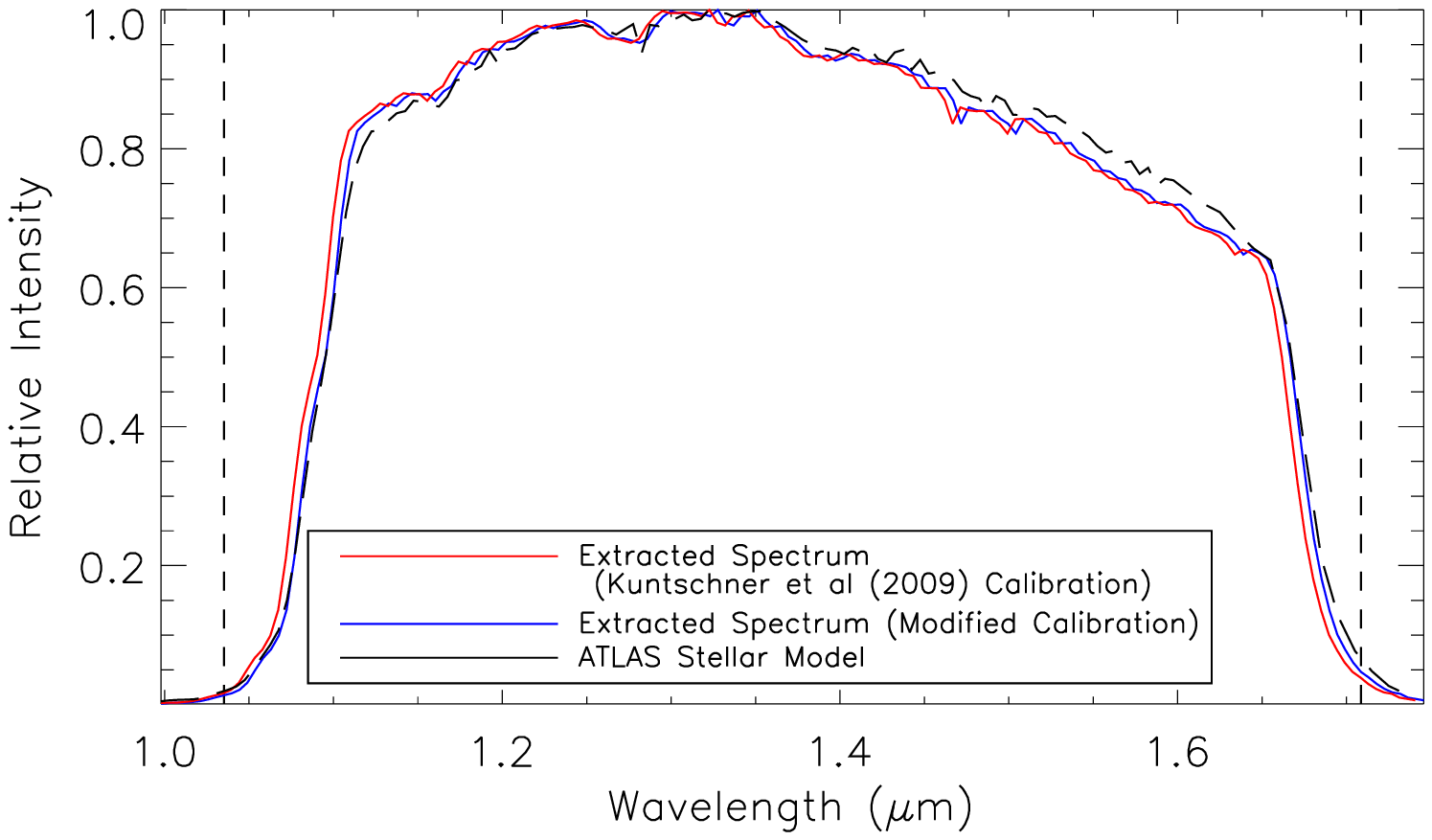}
\caption{Coplotted flat-fielded WFC3 extracted spectra and ATLAS (\citealt{CastelliF.2004}, \citealt{KuruczR.L.1992}) model spectrum approximating TrES-2. The red line is the extracted spectrum calibrated with the wavelength solution described in \citet{Kuntschner2009}. The blue line is the extracted spectrum is calibrated with a wavelength solution shifted by +0.005 $\mu$m. The black line is the ATLAS model spectrum multiplied by the WFC3 sensitivity curve. The spectra are normalized to their highest value. The vertical dashed lines demarcate the edge pixels discarded in analysis (see Section~\ref{sec:manexdata}).\label{fig:treswavecalcorrected}}
\end{figure}

We checked the wavelength solution by comparing an extracted spectrum to a model spectrum. We selected ATLAS stellar models from the published grid\footnote{Available at \url{http://kurucz.harvard.edu/grids.html}} (\citealt{CastelliF.2004}, \citealt{KuruczR.L.1992}) that best matched TrES-2 and WASP-4 and multiplied them by the WFC3 first-order sensitivity curve. Table~\ref{tbl:atlas} gives the grid parameters that define the models selected to approximate these stars. We compared them to the extracted, flat-fielded spectra. We estimate by eye an offset of $0.005~\mu m\approx1.1$ pixel (see Figure~\ref{fig:treswavecalcorrected}), and consequently modified our wavelength solution to $\lambda=(\frac{dl}{dp})_0+(\frac{dl}{dp})_1*l+0.005~\mu m$. Figure~\ref{fig:treswavecalcorrected} presents a comparison between model and observed spectra using this wavelength calibration. We note the inclusion or exclusion of this offset has minimal effect on our conclusions since we ultimately bin the data coarsely in wavelength.

\begin{table}[h]
\begin{center}
\caption{Grid parameters defining ATLAS stellar models used to approximate the target stars. \label{tbl:atlas}}
\begin{tabular}{cccccc}
\tableline\tableline
Star & T$_{\rm{eff}, \star}$ & Z  & log(g) & v$_{turb}$ & Reference\\
 & (K) &  & (cgs) & (km~s$^{-1}$) & \\ 
\tableline
TrES-2 & 5750 & -0.2 & 4.5 & 2.0 &\citet{Kipping2011a}\\
TrES-4 & 6250 & 0.1 & 4.0 & 2.0 &\citet{Chan2011}\\
WASP-4 & 5500 & 0.0 & 4.5 & 2.0 & \citet{Winn2009} \\
CoRoT-1 & 6000 & -0.3 & 4.5 & 2.0 & \citet{Barge2008}\\
\end{tabular}
\end{center}
\end{table}

\section{Light Curve Fitting and Parameter Estimation \label{sec:fitting}}
\begin{figure}[h]
\centering
\includegraphics[width=15 cm, angle=0]{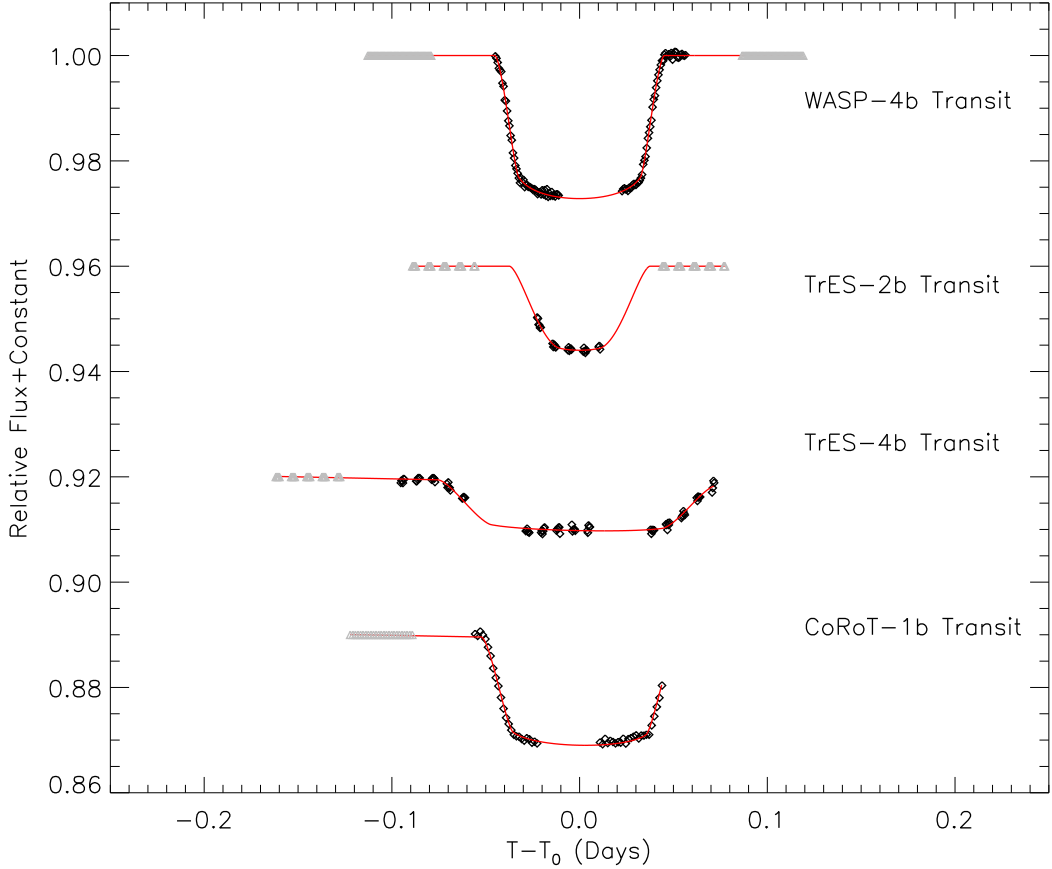}
\caption{Detrended, bandpass-integrated transit light curves for (in order from top) WASP-4b, TrES-2b, TrES-4b, and CoRoT-1b (black diamonds). Plotted in red are the best-fit light curve models. Also plotted are light grey triangles corresponding to the out-of-transit data detrended by themselves (i.e. identically 1). The information from these data have already been incorporated into the in-transit data via detrending, hence they are not included in the fit. The time indicated is $T-T_0$, where $T_0$ is the epoch of transit listed in Table~\ref{tbl:systemparameters}. The flux indicated is the total flux integrated across the bandpass. WASP-4b and TrES-2b are detrended using $divide-oot$, while TrES-4b and CoRoT-1b are analyzed using $single-oot$ (see Section~\ref{sec:singleoot}). \label{fig:corrtransit}}
\end{figure}

\begin{figure}[h]
\centering
\includegraphics[width=15 cm, angle=0]{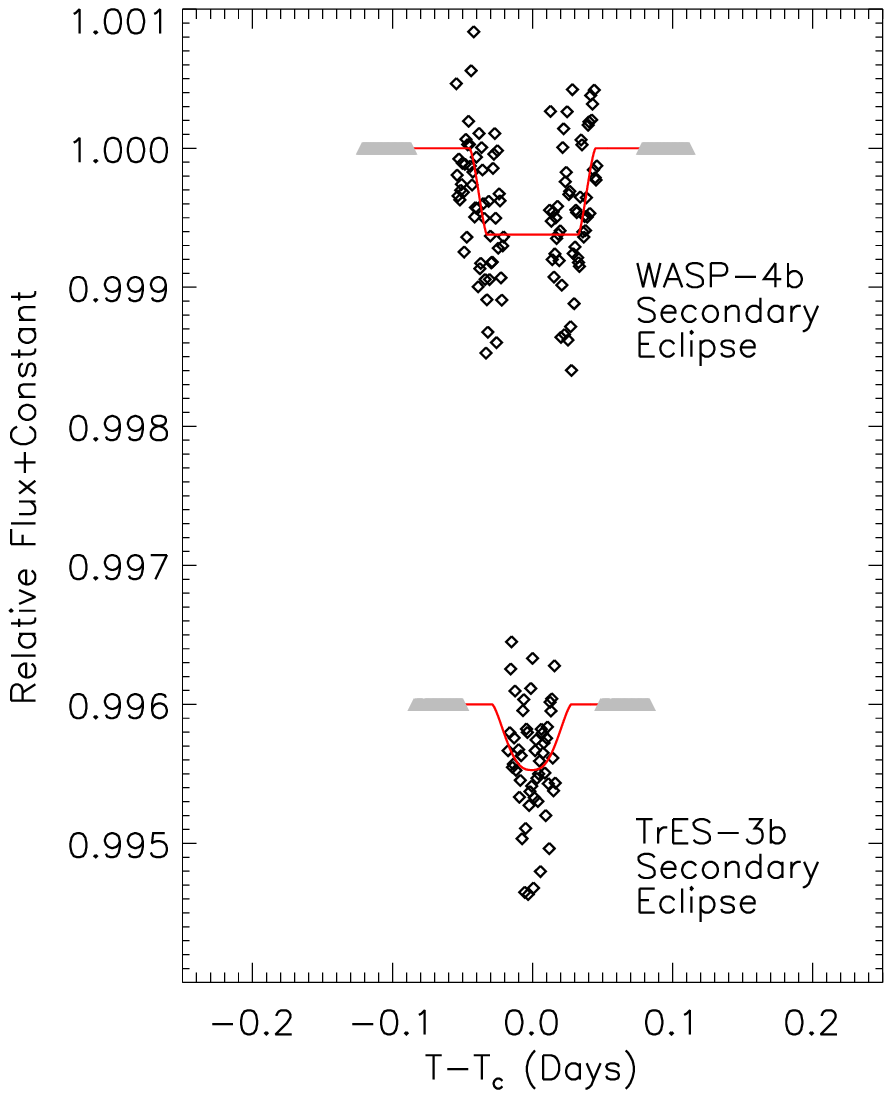}
\caption{Detrended, bandpass-integrated secondary eclipse light curves for WASP-4b and TrES-3b (black diamonds). Plotted in red are the best-fit light curve models. Also plotted are light grey triangles corresponding to the out-of-eclipse data detrended by themselves (i.e. identically 1). The information from these data have already been incorporated into the in-eclipse data via detrending, hence they are not included in the fit. The time indicated is $T-T_c$, where $T_c$ is the epoch of eclipse listed in Table~\ref{tbl:systemparameters}. The flux indicated is the total flux integrated across the bandpass.\label{fig:correclipse}}
\end{figure}

We fit our corrected light curves with transit and eclipse models based on \citet{Mandel2002}.
We use only the in-transit data in our analysis, as the information from the out-of-transit data has already been incorporated into them via the $divide-oot$ data correction: There is no further signal in the out-of-transit data. We fixed the period $P$, the inclination $i$, and the semi-major axis to stellar radius ratio $a/R_{\star}$ from the literature. For transits, we measured the epoch of transit $T_0$ from the white-light event and held it fixed for the analysis of the chromatic light curves. We assumed circular orbits, which is consistent with known constraints\footnote{See \citet{Beerer2011}, \citet{Sanchis-Ojeda2011}, \citet{Kipping2011a}, \citet{Fressin2010}, \citet{Knutson2009}, \citet{Husnoo2012}}. We computed a four-parameter limb-darkening law using the ATLAS stellar models (\citealt{CastelliF.2004}, \citealt{KuruczR.L.1992}) coupled with the WFC3 G141 response curve published by STScI\footnote{http://www.stsci.edu/hst/wfc3/analysis/grism$\_$obs/calibrations/wfc3$\_$g141.html}. Table~\ref{tbl:atlas} lists the parameters describing the ATLAS models we used for each star. The only free parameter was the planet-to-star radius ratio $R_P/R_{\star}$. For secondary eclipses, we assumed an epoch of eclipse $T_c$ based on published transit ephemerides: $T_c=T_0+0.5P$. We fixed $a/R_\star$, $i$, and $R_P/R_{\star}$ to the literature value to establish the eclipse shape. The only free parameter was the eclipse depth. Table~\ref{tbl:systemparameters} summarizes the system parameters derived or adopted in the course of this study. Figures~\ref{fig:corrtransit} and ~\ref{fig:correclipse} show the detrended white-light transits and secondary eclipses with best-fit model light curves coplotted.

\begin{table}[h]
\footnotesize
\caption{System parameters assumed in single-parameter light curve fits\label{tbl:systemparameters}}
\begin{tabular}{ccccccc}
\tableline\tableline
Event & P  & i &  $a/R_{\star}$ & $R_P/R_{\star}$ &$T_0$ or $T_c$  & References \\ 
 & (days)  & ($^\circ$) & &  & (BJD-2455000) &  \\ 
\tableline
TrES-2 Transit & 2.47061896 & 83.952 & 7.9830& -- & 479.5333$\pm0.0002$ &  \citet{Kipping2011a}\\
TrES-3 Eclipse & 1.30618608 & 81.99  & 6.02 & 0.1661 & 623.36823 &  \citet{Christiansen2011}\\
TrES-4 Transit & 3.5539268 & 82.81 & 6.08 & -- & 524.5360$^{+0.0008}_{-0.0007}$ &  \citet{Chan2011}\\
WASP-4 Transit & 1.33823187 & 88.80 & 5.482 & -- & 526.16356$^{+0.00007}_{-0.00008}$ &\citet{Sanchis-Ojeda2011}\\
WASP-4 Eclipse & 1.33823187 & 88.80 & 5.482 & 0.156 & 528.17110 &  \citet{Sanchis-Ojeda2011}\\
CoRoT-1 Transit & 1.5089656 & 83.88 & 4.751 & -- & 950.5993$^{+0.0002}_{-0.003}$ & \citet{Bean2009}\\
\tableline
\end{tabular}
\end{table}

We fit each wavelength channel of data for each object individually using least-squares (LS), Markov-Chain Monte Carlo (MCMC), and Residual Permutations (RP) analyses. A least-squares analysis assumes the errors on the data are distributed in a Gaussian fashion. We used the least squares method to provide an initial fit to the data. We used the nonlinear least-squares IDL fitting package MPFIT \citep{Markwardt2009}, a robust implementation of the Levenberg-Marquardt algorithm, to fit the data and estimate parameters. We measure the resulting reduced chi-squared $\chi^2$. We have so far assumed purely photon noise. Were our errors purely photon noise and our model an accurate description of the data, we would observe $\chi^{2}=N$, where $N$ is the number of degrees of freedom. Instead, we generally observe $\chi^{2}>N$. We assume our model is an accurate description of the data, in which case $\chi^{2}>N$ implies the presence of additional noise sources. To account for this, we rescale the errors on our data by a factor $\sqrt{\chi^{2}/N}$ and rerun the least squares fit. Table~\ref{tbl:ppm} gives the precision of each channel based on the standard deviation of the residuals in parts per million (PPM), compared to the expected photon error. The mean scatter in the residuals ranges from 5$\%$ to 60$\%$ above the photon noise. For comparison, \citet{Berta2011} derives a scatter in the residuals 10$\%$ above photon noise.

\begin{table}[h]
\footnotesize
\begin{center}
\caption{PPM precision of fit for each channel for each objects.\label{tbl:ppm}}
\begin{tabular}{p{1.5 cm}p{1.4 cm}p{0.9 cm}p{0.9 cm}p{0.9 cm}p{0.9 cm}p{0.9 cm}p{0.9 cm}p{0.9 cm}p{0.9 cm}p{0.9 cm}p{0.9 cm}p{0.9 cm}}
\tableline\tableline
Event &  Photon Noise & Mean &Bin 0 &  Bin 1 &  Bin 2 &  Bin 3 &  Bin 4 &  Bin 5 &  Bin 6 &  Bin 7 &  Bin 8 &  Bin 9 \\
 &  (PPM) & (PPM) & (PPM) & (PPM)  &  (PPM)  & (PPM)  &  (PPM)  &  (PPM)  & (PPM)  &  (PPM)  &  (PPM)  &  (PPM)  \\
\tableline
TrES-2 Transit &  975 & 1060 & 869   &    859   &   1060    &   1010   &    1080   &   1100    &   1150    &   1180  &     1060  &     1280\\
TrES-3 Eclipse &   838 & 1340 & 1770   &   1440   &    1240   &    1410   &    1700   &    1470   &    1220  &     905  &     1200   &    1040\\
TrES-4 Transit &  1270 & 1460 & 1520   &    1400  &     1410   &    1870  &     1140   &    1340   &    1380   &    1180    &   1500   &    1830\\
WASP-4 Transit &    914 &  1260 & 1140 & 1390 & 1570 &  1230 & 1270 & 1350 & 1260 & 1340 & 1060 &  972\\
WASP-4 Eclipse &   914 & 1210 & 1060   &    1300 &       1200 &      1460 &       1250 &  1330 &       1110 &       1230 &       1070 &       1080\\
CoRoT-1 Transit &    1110 & 1160 & 1310    &   1110  &     1040    &   1290    &   1220 &  1340  &     1020  &    1110   &    1200   &    1020\\
\tableline
\end{tabular}
\tablecomments{Bin 0 is the blue-most and bin 9 the red-most. The photon noise limit for each bin is given in the second column. Also given is the mean per-channel precision for each dataset.}
\end{center}
\end{table}

\subsection{MCMC Analysis}
An MCMC analysis, while still assuming the errors to be Gaussian, allows for a more thorough exploration of parameter space than a least squares analysis and accounts for degeneracies between parameters. For MCMC fitting, we used the Metropolis-Hastings algorithm with the Gibbs sampler (see \citealt{Ford2005}). The MCMC chains were run to a length of $3\times10^5$ links. The first $1/3$ of these were eliminated to exclude "burn-in", leaving $2\times10^5$ links for parameter estimation. We verified that the chain contained many autocorrelation lengths (at least 30), implying our chain contained many independent realizations of the posterior. We checked the shape of our posterior distributions. We found them to be centrally peaked. We took the median (50th percentile value) of the distribution to estimate the parameter values, and the the 16th and 84th percentile values of the distribution to mark the extent of the 1-$\sigma$ error bars.

\subsection{Residual Permutations Analysis}
We also consider the residual permutations (RP) method. A residual permutations analysis is sensitive to time-correlated noise. Our residual permutations analysis was similar to the treatment in \citet{Beerer2011}. We subtracted off the least squares fit from the data to generate a chain of residuals. We permuted each element in that chain by 1, added it back to the model to generate a synthetic dataset, and refit the data with least squares. We iterated this procedure until each element on the chain had cycled all the way around, keeping track of the parameter estimates as we went. The histogram of the best-fit values forms an approximation to the posterior parameter probability density function, and its width is another estimate of the parameter error. We took the median (50th percentile value) of the distribution to estimate the parameter values, and the the 16th and 84th percentile values of the distribution to mark the extent of the 1-$\sigma$ error bars.

\begin{figure}[h]
\centering
\includegraphics[width=15 cm, angle=0]{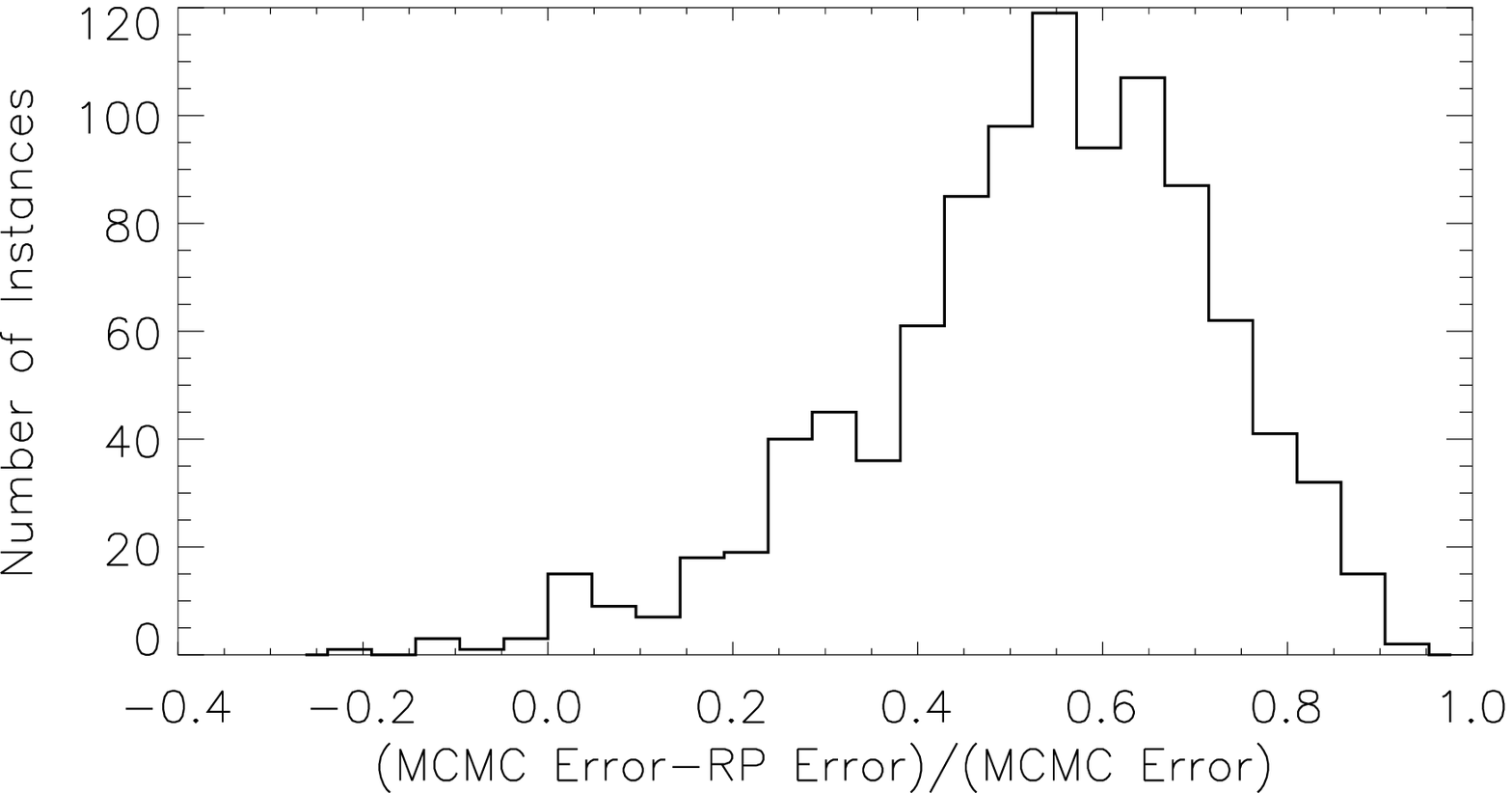}
\caption{Comparison of the performance between MCMC and RP. The histogram shows the fractional differences $(\sigma_{MCMC}-\sigma_{RP})/\sigma_{MCMC}$ between the residual permutation error estimate $\sigma_{RP}$ and the MCMC error estimate $\sigma_{MCMC}$ for a suite of 1000 simulated WASP-4b eclipses (Gaussian noise only). Typically $\sigma_{RP}<\sigma_{MCMC}$. For the WASP-4b eclipse dataset, the residual permutations analysis tends to underestimate the error. \label{fig:fakedataspreads}}
\end{figure}

We tested our MCMC and residual permutations analysis on simulated WASP-4b eclipse data. We produced a simulated WASP-4b secondary eclipse light curve using the observed phases and assuming Gaussian noise. We found that the residual permutations method tended to underestimate the parameter error. To understand why, we remember that for the $divide-oot$ analysis all the data fitted is in-transit or in-eclipse. Consider the limiting case of an observation with all the data in-eclipse and equal errors on each data point.  No matter how much one permutes the residuals and adds them back in, the synthetic dataset ends up as a permutation of the original data, resulting in repeated identical widths and a 0-width posterior. This scenario was borne out in simulations. We simulated 1000 WASP-4b eclipses with identical eclipse parameters and Gaussian noise levels, measured the mean and 1-$\sigma$ widths for the posterior distribution, and tested how well they estimated the dataset characteristics. Figure~\ref{fig:fakedataspreads} summarizes the results of this simulation. The results vary, but typically the residual permutations method underestimated the error on eclipse depth by 60$\%$.

The MCMC may  understimate error in the limit of high time-correlated (red) noise, while the residual permutations technique is sensitive to red noise on timescales different from the transit/eclipse duration \citep{Southworth2008}. We therefore adopt the maximal error estimate to be the parameter error: i.e., we assign the error on eclipse depth $R_P/R_{\star}$ to be the maximum 1-$\sigma$ error obtained from the MCMC, residual permutations, and least-squares methods. We must adopt the maximal estimate as different methods account for different sources of error. We found that while the MCMC and least squares error estimates were in close agreement (as expected for 1-parameter fits), the residual permutations error estimates were sometimes divergent. Of our datasets, we found the CoRoT-1b transit to be dominated by residual permutations errors, indicating the presence of correlated noise. Residual permutations and MCMC errors were comparable for WASP-4b (eclipse and transit) and the TrES-4b transit, while the MCMC errors dominated TrES-2b and TrES-3b. 

\subsection{Fitting Transit Light Curves with Reduced Baselines: TrES-4 and CoRoT-1\label{sec:singleoot}}
TrES-4 and CoRoT-1 pose a challenge for our method. For these objects, only one out-of-transit orbit is available (aside from the very first, discarded, settling orbit). For these objects, we adapt our method as follows: first, we divide our in-event data by the single stable out-of-transit orbit available. This removes periodic effects, such as the ramp. However, we additionally observe a visit-wide linear trend in flux that varies by channel. We observe this remnant trend in CoRoT-1 and TrES-4, as well as TrES-2 and WASP-4. $divide-oot$ naturally corrects this effect since it interpolates in flux between the first and last out-of-transit orbit. Since we lack a second out-of-transit orbit, we must fit out this trend.  We multiply our transit model by a linear trend and fit for three parameters simultaneously: the slope and intercept of the systematic trend, and $R_P/R_{\star}$. We note that the linear trend is an important part of this detrending, as omitting it from the analysis leads to spurious features appearing in the spectrum. We term this method the $single-oot$ method.

No correlations are evident in the residuals of the TrES-4b transit dataset studied with $single-oot$. However, when the transit of CoRoT-1b is studied in integrated white light, a faint linear trend as a function of HST orbital phase is observed in the residuals. This trend weakens when the dataset is decomposed into multiple channels. In most channels, it does not appear, while in a few it is faintly visible. In no channel is it as clearly visible as in the bandpass-integrated channel. This suggests an instrumental effect that is not completely corrected by the single-oot method, and that is wavelength sensitive (since it fades in higher resolution observations). We note that the channels that show hints of this trend also show inflated residual permutations error bars, indicating our code is accounting for this correlated noise.

\begin{figure}[h]
\centering
\includegraphics[width=15 cm, angle=0]{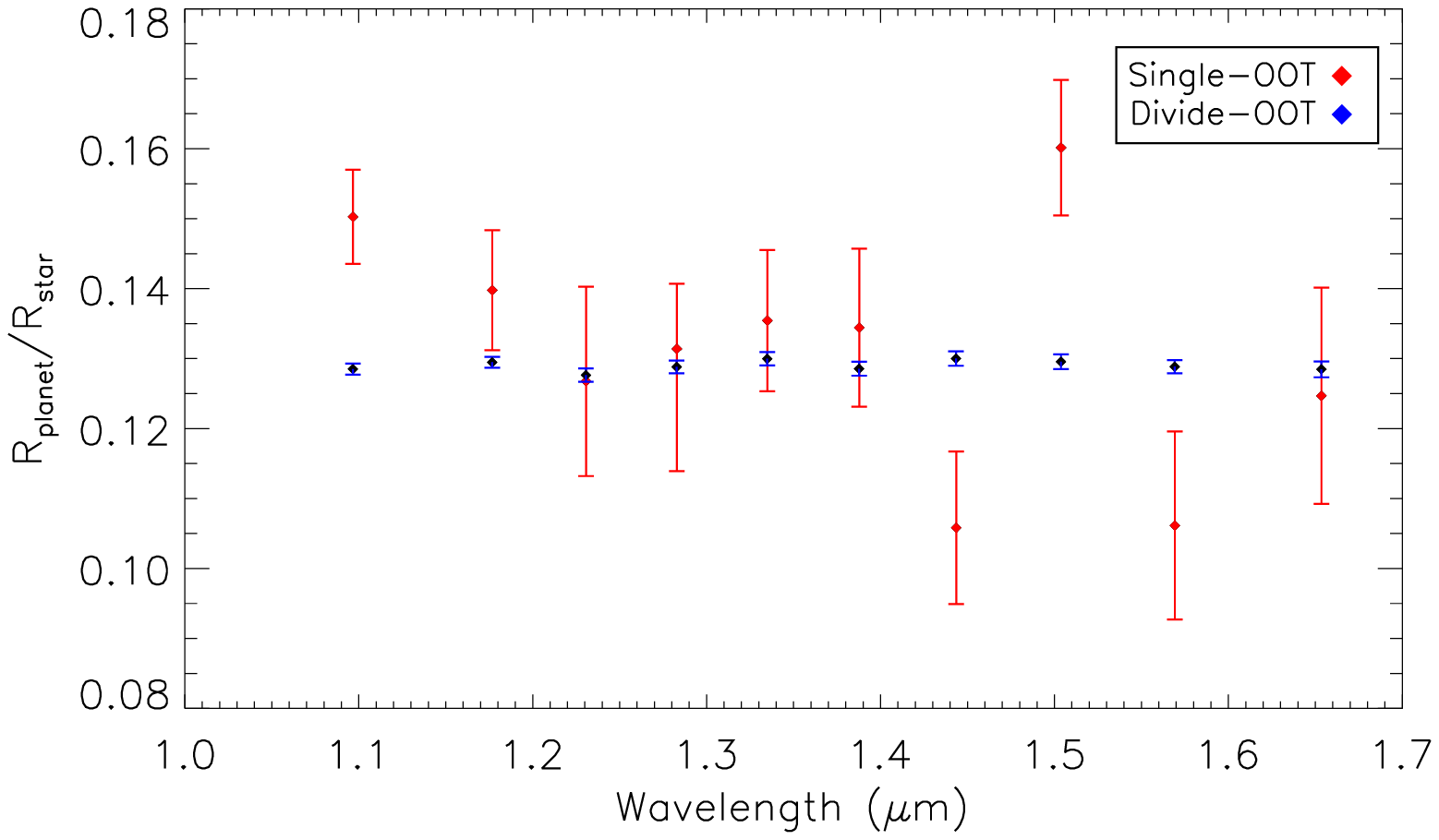}
\caption{Results of test evaluating impact of a reduced baseline on precision per wavelength element of resolution. This plot presents transmission spectra of TrES-2b computed using the $divide-oot$ (blue) and $single-oot$ (red) methods. The two spectra are consistent in overall eclipse level and spectral shape. The errors are much higher for $single-oot$. Based on this study, we expect our TrES-4b and CoRoT-1b spectra to be low-precision. \label{fig:linnonlinetres2}}
\end{figure}

Unlike the $divide-oot$ method, the $single-oot$ method parametrizes some of the systematics. To validate the $single-oot$ method, we apply it to the TrES-2 dataset and test for consistency with the $divide-oot$ method. We discard the second OOT orbit of TrES-2, use only one OOT orbit to correct the data, and fit the transit with the three-parameter model described above. We compute single-channel and ten-channel-spectra for TrES-2b. With more parameters to trade off against, the error on $R_P/R_{\star}$ increases by a factor of 10. Figure~\ref{fig:linnonlinetres2} coplots the $single-oot$ and $divide-oot$ spectra. The two spectra are consistent in overall depth and contain no statistically significant features. However, the spectra derived using $single-oot$ are much less precise. This is not unexpected, as the model has more parameters that trade off error against each other. $single-oot$ performs similarly with the WASP-4b transit dataset. We conclude that  $single-oot$ and $divide-oot$ yield consistent results, but $single-oot$ is significantly less precise than $divide-oot$. 

\section{Diagnostics and Checks\label{sec:validation}}
In this section, we describe tests we performed on our data to test the robustness of our analysis.

\subsection{Residuals\label{sec:residuals}}
To assess the quality of our fits, we studied the residuals of our fits on a channel-by-channel basis. We visually checked 1) for evidence of correlations in the residuals, 2) confirmed that the histogram of the residuals was centrally peaked, and 3) rebinned the residuals and measured their variance to look for evidence of red noise as described in \citet{Pont2006}. We do not observe correlations in the residuals and confirm them as centrally peaked. We found the RMS of the rebinned residuals to fall off as $1/\sqrt{N}$, where N is the number of residuals in each bin; this indicated low red noise. In the few cases where the RMS of the rebinned residuals fell off slower than $1/\sqrt{N}$, the error bar inflated relative to the other bins, indicating our code was accounting for the red noise.

\subsection{Allowing Limb-Darkening to Float\label{sec:LDfloat}}
We considered the possibility that fixing the limb-darkening to theoretically computed values might introduce spurious features, as limb-darkening could be partially degenerate with transit depth. To test this hypothesis, we analyzed the WASP-4b transit dataset while letting the limb-darkening float. We adopted a quadratic limb-darkening law as described in \citet{Winn2009}, and fitted jointly for $R_P/R_{\star}, u_1$, and $u_2$, subject to the constraint that $0<u_1+u_2<1$, which ensures that the limb will be dimmer than the center. We started our MCMC and least-squares fitting at the theoretical expectation for $u_1$ and $u_2$. The resulting spectrum was consistent to better than $1-\sigma$ with the fixed limb-darkening analysis, indicating that fixing limb-darkening was not introducing spurious features. Allowing the limb-darkening to float degraded spectral precision by $10\%$.

\subsection{Effect of Flat-Fielding}
Our data are not flat-fielded. Since we are conducting a differential measurement and our drift is low (hundredths of a pixel), we do not need to flat field. However, as a test, we examine transmission spectra for TrES-2b and WASP-4b with and without flat-fielding. The wavelength solution is an input to determining the flat field, so we compute the flat-field with and without our additional $0.005 \mu$m offset. These test objects were chosen because the TrES-2b spectra were close to photon noise while the WASP-4b spectra were farther from it. We constructed a color-dependent flat-field as described in the aXe User Manual\footnote{\url{http://axe.stsci.edu/axe/manual/html/index.html}} and flat-fielded each frame, and then let our analysis proceed as described in Section~\ref{sec:fitting}. We find the derived spectra deviate from our previous results by less than $1-\sigma$. This test confirms flat-fielding does not affect our analysis.

\subsection{Other Potential Systematics}
We checked for confounding systematic effects. For each channel, we visually monitored the background and flux and looked for evidence of an anomaly uncorrected by the $divide-oot$ method. We found no such anomalies. We monitored all HST engineering and photometry keywords in the \emph{flt} file headers for evidence of deleterious correlations due to variations in the instrument state. None were found. We tested whether $divide-oot$ corrected out the effects of the 2D motion (drift) of the spectrum on the detector. For the WASP-4b secondary eclipse, we measured the 2D drift of the spectrum on the detector over the course of observations. We plotted the drift against the un-detrended fluxes and observed a correlation. We then plotted drift against residuals and observed no correlation, indicating our analysis corrects out the influence of drift. 

\subsection{Recovery of Previous GJ1214b Results}
As a check of our methods, we ran our pipeline on Visit 3 of the GJ1214 transit data collected by \citet{Berta2011}. We ran it on their \emph{flt} files, allowing us to validate both our spectral extraction routines as well as our detrending and fitting routines. We analyzed the white light curve of  this data, allowing  $R_P/R_{\star}$, $a/R_{\star}$, $i$, and $\Delta T_0$ to float. We used the two-parameter limb-darkening law computed by   \citet{Berta2011} from PHOENIX models, and used their ephemeris as a starting point for the fit.  Our parameter estimates and ephemeris are consistent with \citet{Berta2011} within $1-\sigma$.

\begin{figure}[h]
\centering
\includegraphics[width=15 cm, angle=0]{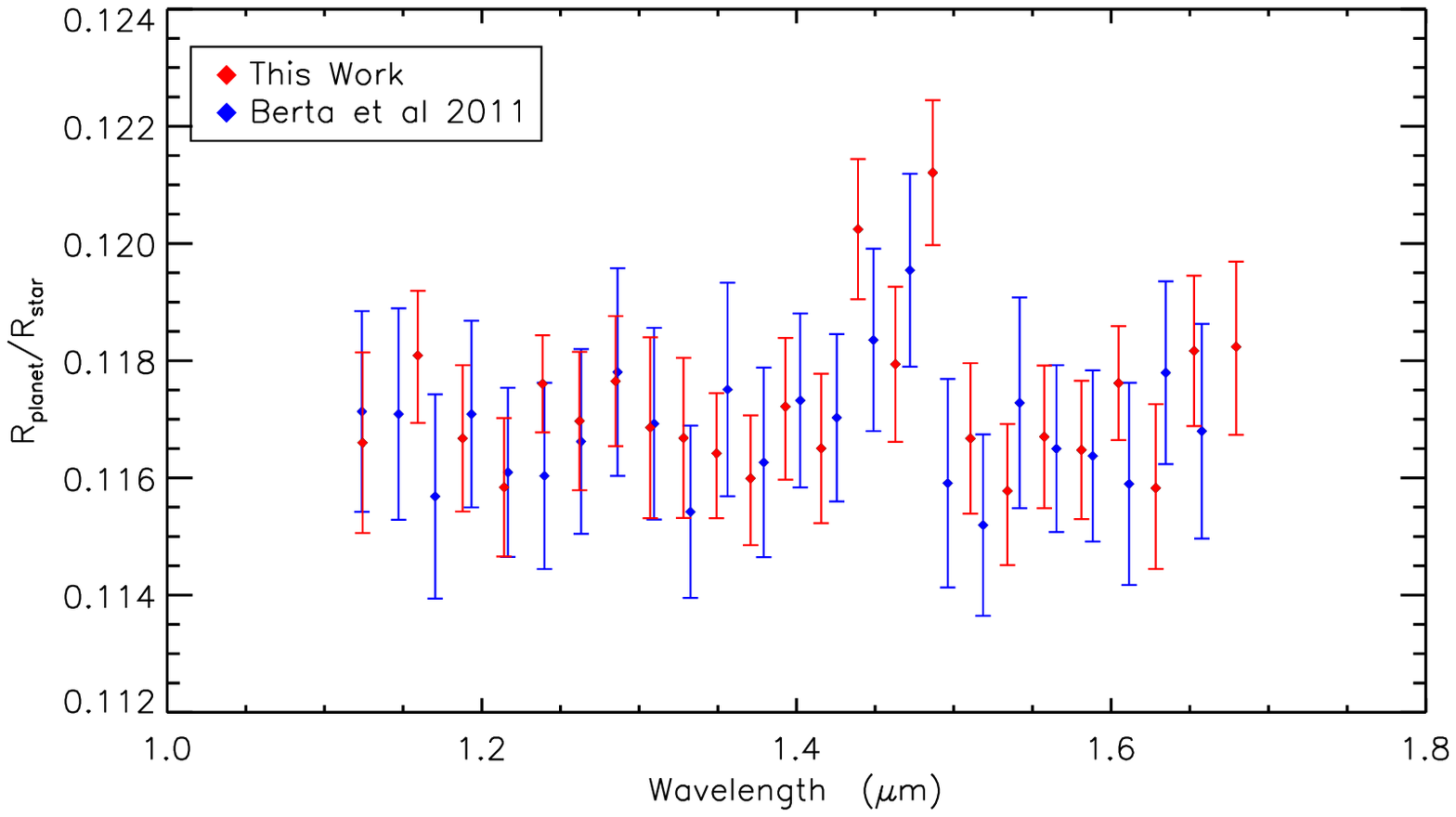}
\caption{Transmission spectra of GJ1214b extracted from Visit 3 of the WFC3 data analyzed in   \citet{Berta2011} The blue data are derived by the \citet{Berta2011} pipeline, the red data were derived from ours. \citet{Berta2011} compute 2-parameter square-root limb-darkening laws from the PHOENIX models and use them as priors in the fit. We use the same PHOENIX models (provided courtesy of Z. Berta) to compute the corresponding 4-parameter nonlinear limb-darkening, and fix these laws in the fit.  \label{fig:bertaranjan}} 
\end{figure}

We next moved to compute spectra by breaking the white-light curve into channels, fixing all parameters other than $R_P/R_{\star}$. We adopted the same transit parameters used in \citet{Berta2011}, and used code kindly provided by them to compute 4-parameter nonlinear limb-darkening laws from the PHOENIX models. The resulting spectra are coplotted with the earlier published results in Figure~\ref{fig:bertaranjan}. We find that the two are consistent within $1-\sigma$. Our error bars are slightly smaller because we assume a limb-darkening law, while they use their computed laws as priors. As an additional test, we also ran a $single-oot$ analysis on this GJ1214 dataset. As expected, we find a flat spectrum consistent with the level of the  \citet{Berta2011} spectrum, but with precision degraded by an order of magnitude.

\subsection{Test of $differential-eclipse$\label{sec:diffec}}
We also explored an alternate method of systematics decorrelation, based on but not identical to techniques being developed by Drake Deming and Ashlee Wilkinson \citep{Deming2013}. We term this method $differential-eclipse$. In this method, each extracted spectrum is normalized by the total flux in the spectrum. This removes achromatic variations in flux, such as the ramp, from the light curve. The transit is also removed. However, \emph{chromatic} variations remain: during eclipse, wavebands corresponding to planetary opacity will show reduced flux, while wavebands free of planetary opacity will show correspondingly enhanced flux. This method produces a differential spectrum: it can be used to set the shape of the spectrum but not its overall level. We applied this method to TrES-2b and WASP-4b in transit. We fixed all parameters except $R_P/R_{\star}$ to literature values. We fixed $R_P/R_{\star}$ to the level extracted from our $divide-oot$ white-light analysis. To fit the data, we found it necessary to decorrelate against an overall linear trend, and the x- and y-drifts\footnote{The 2D motion of the spectrum on the detector over time compared to the original position.}. We measured the x-drift by determining the peak of the cross-correlation function between the first extracted spectrum and subsequent spectra.  To measure the y-drift, we summed each image over the x-dimension yielding a vector showing the distribution of flux received as a function of y-pixel. We measured the cross-correlation function between the vector corresponding to the first image and subsequent images.\footnote{We experimented with creating templates based on the mean of all the spectra, but obtained equivalent results.} Our final 5-parameter model took form $F(t)=(1-a_0*T(t))(a_1+a_2t)(1+a_3*x_d)(1+a_4*y_d)$, where $F$ is flux, $t$ is time, $T(t)$ represents the transit shape, $x_d$ represents the x-drift, and $y_d$ represents the y-drift\footnote{\citet{Deming2013} account for shifts in the spectra prior to fitting and hence fit for just the linear baseline and transit depth.}. We verified that the resulting spectra were zero mean as expected. We fit the data and estimated the errors on a channel-by-channel basis as described for the $divide-oot$ method, rejecting outliers, inflating errors, and choosing the maximum errors derived from least-squares, MCMC, and residual permutations analyses. The resulting spectra are consistent and have similar precision.  As we do not show increased performance from this method, we elect to continue to use $divide-oot$ in our analysis. 

\subsection{Test of $differential-oot$\label{sec:diffoot}}
We tried to capture the best of both $differential-eclipse$ and $divide-oot$ by hybridizing them into what we termed the $differential-oot$ method. We detrended achromatic effects by normalizing extracted spectra by the total flux in each spectrum as described in Section~\ref{sec:diffec}. We then implemented $divide-oot$ to correct out additional chromatic effects. We fit each channel with a transit model with transit parameters set from literature to establish the transit shape, with the free parameter being transit depth. We derived the resultant spectrum for WASP-4  and compared it to the spectrum derived using $divide-oot$. The two spectra are consistent and have similar precisions. As we did not gain in precision by using this technique, we again elected to continue using $divide-oot$ for our analysis.

\subsection{Treatment of Saturated Reads}
\begin{figure}[h]
\centering
\includegraphics[width=15 cm, angle=0]{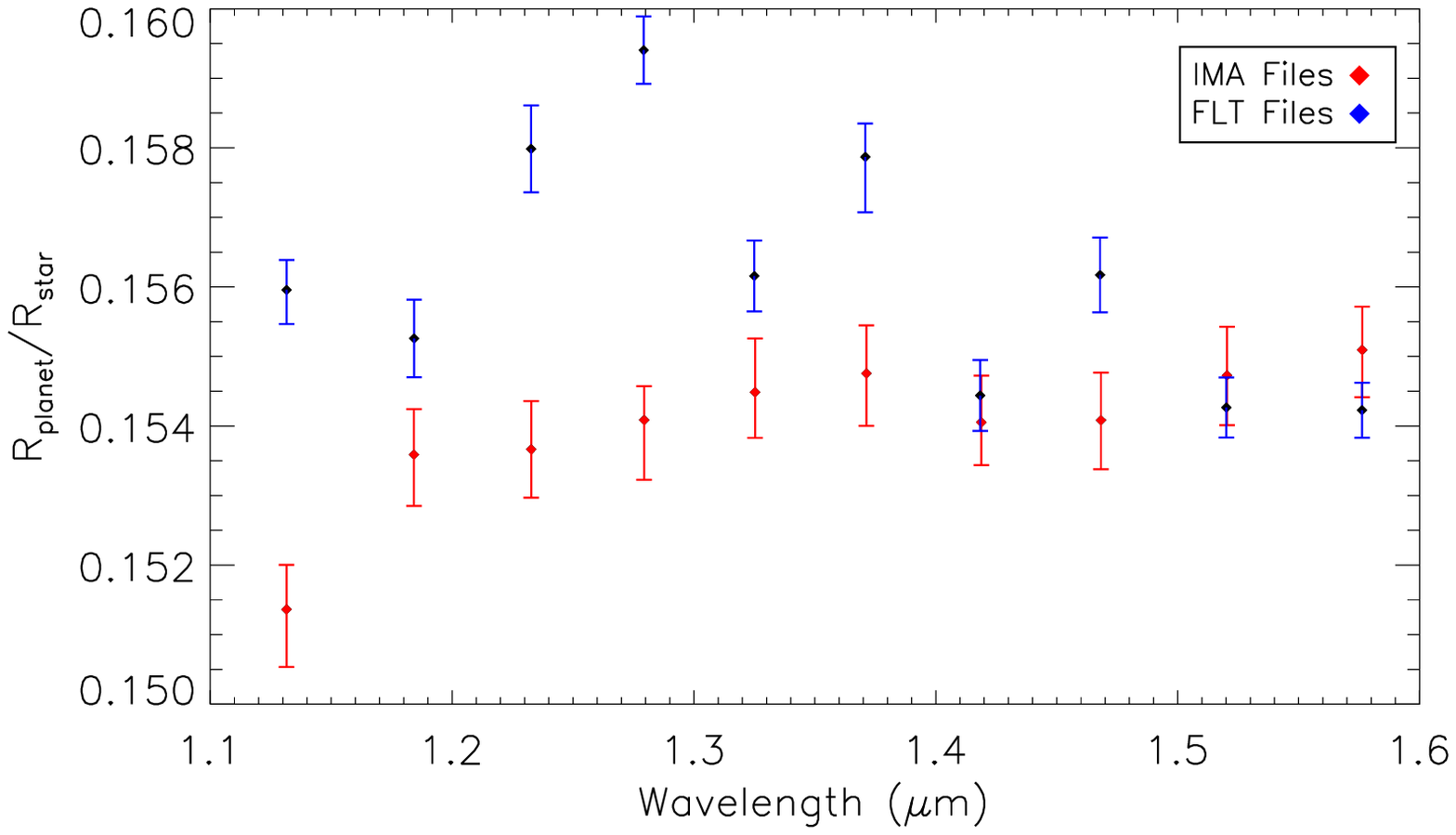}
\caption{Transmission spectrum of WASP-4b derived using different treatments of saturated reads. In blue are the data derived from the WFC3 \emph{calwf3} pipeline (the \emph{flt} files). In red are the data derived from a simple linear fit to the individual nondestructive reads, using only the first 3/6 reads (excluding zeroth read) to assure that the data are not saturated. The spectra are inconsistent. The other datasets are not affected by this problem.\label{fig:sattest}}
\end{figure}

Some of our data are exposed to levels exceeding nonlinearity and/or saturation. We rely on the \emph{calwf} pipeline to flag and reject saturated reads when estimating flux on a pixel-by-pixel basis. A pixel is considered saturated if its response deviates by more than $5\%$ from linearity. Until this point, the pixel is corrected for nonlinearity via a polynomial correction; after this point, the pixel is considered saturated and future reads are not used to estimate count rates. This point occurs at roughly 78,000 electrons\citep{Dressel2012}. To test the effectiveness of this pipeline, we compared the spectra derived using the \emph{calwf3}-corrected \emph{flt} files to those derived from the individual calibrated reads stored in the \emph{ima} files. We determine the count rates from the \emph{ima} files by fitting a linear trend to the counts measured in each read as a function of time on a pixel-by-pixel basis; the fit weights the data by the Poisson error and an assumed read noise of 20.4 electrons\citep{Dressel2012}, added in quadrature. In these fits, excluding the zeroth read, we included only the first 3/6 reads for the WASP-4b and TrES-3b data, 9/15 reads for TrES-2b, 10/15 reads for TrES-4b, and 8/15 reads for CoRoT-1b. This corresponds to a limit of $\approx$ 40,000 accumulated electrons per pixel in the extracted spectrum, well short of saturation level. We then derive spectra from the data as described in Sections~\ref{sec:extract} and ~\ref{sec:fitting}. We find the spectra derived from the \emph{flt} and \emph{ima} files to be consistent for every dataset, except the WASP-4b transit. As Figure~\ref{fig:sattest} shows, the two spectra derived for WASP-4b are inconsistent: the spectrum derived from the \emph{flt} files shows enhanced absorption absent from the spectrum derived from the \emph{ima} files. The region showing this absorption corresponds to the most heavily exposed pixels. Since we have no reason to favor either approach, we conclude that the WASP-4b transmission spectrum is sensitive to treatment of saturated reads and hence nonrobust, and consequently do not report a spectrum for it.

\section{Results and Discussion\label{sec:prelimres}}
In this section, we present our WFC3 transmission and emission spectra for the planets in our sample and compare them to model spectra in order to constrain the atmospheric properties of the planets. We model the atmospheric spectra for all the planets using the exoplanetary atmospheric modeling and retrieval technique of \citet{Madhusudhan2009b, Madhusudhan2010a}. For each dataset in our sample, we explore the space of atmospheric chemical composition and temperature structure to identify models that explain the data. For several cases in the present work, however, the  observational uncertainties in our spectra allow multiple model solutions to the data, in which case we seek to identify the generic families of models (e.g. carbon-rich versus oxygen-rich; \citealt{Madhusudhan2012}) that explain the data. In the WFC3 bandpass (1.1-1.7 $\mu$m), the dominant opacity in the O-rich models is due to H$_2$O, whereas the dominant contributions in C-rich models arise from the hydrocarbons (CH$_4$, HCN, and C$_2$H$_2$). In the cases of WASP-4b and TrES-3b in thermal emission, where previously published observations are available in other spectral bandpasses, we combine our data with published data to further refine our model constraints whereby contributions to opacity due to other molecules (e.g. CO and CO$_2$) also become relevant.

The model computes line-by-line radiative transfer in a plane-parallel atmosphere in local thermodynamic equilibrium (LTE), and assumes hydrostatic equilibrium and global energy balance. The model atmosphere includes the major sources of opacity expected in H$_2$-dominated atmospheres, over a wide range of temperatures and carbon-to-oxygen (C/O) ratios, adopted from \citet{Madhusudhan2012}. In addition to the H$_{2}$O absorption expected to dominate a solar-composition (low C/O ratio) atmosphere, the opacity sources considered include CO, CH$_{4}$, CO$_{2}$, NH$_3$, C$_2$H$_2$, and HCN, and continuum opacity due to H$_{2}$-H$_{2}$ collision-induced absorption (CIA). The molecular line-lists for H$_2$O, CO, CH$_4$, and NH$_3$ in our model were obtained from \citet{Freedman2008} and references therein. Our line-lists for CO$_2$,  C$_2$H$_2$, and HCN were obtained from \citet{Wattson1986}, \citet{Rothman2005}, and \citet{Harris2008}, respectively. Our CIA opacities were obtained from \citet{Borysow1997} and \citet{Borysow2002}. The volume mixing ratios of the molecules, i.e. the chemical composition, and the pressure-temperature ($P$-$T$) profile of the 1-D atmosphere are input parameters to the model \citep{Madhusudhan2009b, Madhusudhan2012}. 

Given the planetary properties and the parametric temperature profile and molecular abundances, the model computes a spectrum for the required geometry; a  transmission spectrum at primary eclipse or a thermal emission spectrum at secondary eclipse (see e.g. \citealt{Madhusudhan2009b}). While a transmission spectrum probes the atmosphere at the day-night terminator of the planet, a thermal emission spectrum probes the dayside atmosphere of the planet as observed at secondary eclipse. In order to compute the planet-to-star flux ratios in thermal emission, we assume an ATLAS model for the stellar spectrum \citep{CastelliF.2004}. For TrES-3b, we use a model with T$_{\rm{eff}}$=5750 K, Z=0.0, and log(g)=4.5. For WASP-4, we use a model with the same parameters specified in Table~\ref{tbl:atlas}.

 \subsection{Transmission Spectra}
\begin{figure}[h]
\centering
\includegraphics[width= 6 in, angle=0]{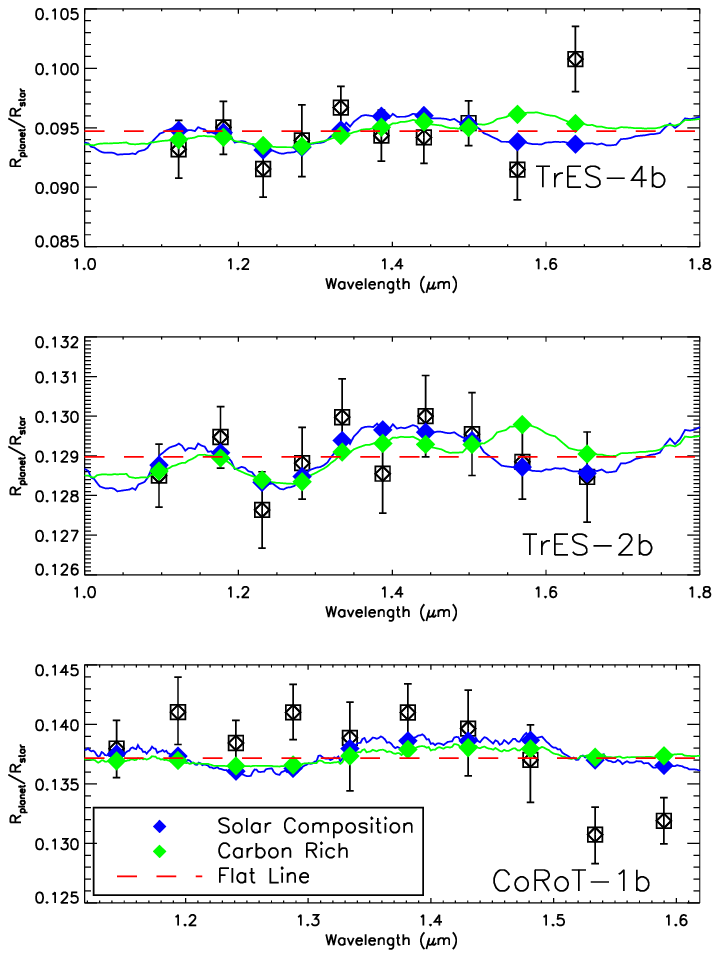}
\caption{10-channel transmission spectra for (from top) TrES-4b, TrES-2b, and CoRoT-1b. Coplotted are best-fit atmospheric models. In blue is a standard solar-composition atmosphere. In green is a carbon-rich atmosphere. The colored diamonds show the bandpass-integrated models. Also plotted is best-fit flat line, in red. \label{fig:combinedtransmissionspectra}}
\end{figure}

 Figure~\ref{fig:combinedtransmissionspectra} presents the transmission spectra derived in this paper using $divide-oot$ (TrES-2b) and $single-oot$ (TrES-4b, CoRoT-1b), compared to representative cloud-free solar-composition and carbon-rich atmospheric models, as well as a flat line. Table~\ref{tbl:tspec} presents the spectra in tabular form. We allowed a constant offset in $R_P/R_{\star}$ when fitting the models, hence there are 9 degrees of freedom associated with these 10-channel spectra. Table~\ref{tbl:tchisq} gives the $\chi^{2}$ values of the fits of these models to the data,  the average uncertainty in $R_P/R_\star$ per wavelength element $\sigma_{R_P/R_{\star}}$, and the corresponding uncertainty in the transit depth, $\sigma_d=2(R_P/R_\star)\sigma_{R_P/R_{\star}}$. It also gives the quantity $\delta=2HR_P/R_\star^2$. This quantity is  the ratio of the cross-sectional area of an annulus of width $H$ of the planet's atmosphere to the surface area of the star, and as such gives the scale of the signal expected from the planet's atmosphere \citep{Brown2001}. We may take the typical atmosphere to have height $\sim5 H $\citep{SeagerBook}; hence, a spectral feature that is optically thick across the planetary atmosphere can be expected to produce a transmission signal on the scale of $5\delta$. 

\begin{table}[h]
\footnotesize
\begin{center}
\caption{Transmission spectroscopy results. \label{tbl:tchisq}}
\begin{tabular}{ccccccc}
\tableline\tableline
Planet & $\chi^{2}$& $\chi^{2}$  & $\chi^{2}$ & $\sigma_{R_P/R_{\star}}$ &$\sigma_{d}$& $\delta$\\ 
& (Solar-Composition) & (Carbon-Rich) & (Flat Line) & & ($2(R_P/R_\star)\sigma_{R_P/R_{\star}}$)& ($2HR_P/R_\star^2$)\\
\tableline
TrES-2b & 2.84 & 4.57 & 5.49 & 0.00095 & 0.00025&  0.00012\\
TrES-4b & 11.2 & 10.7 & 10.5 & 0.0023 & 0.00045& 0.00016\\
CoRoT-1b & 21.6 & 24.4 & 23.0 & 0.0026 & 0.00076& 0.00025\\
\tableline
\end{tabular}
\end{center}
\end{table}

\begin{table}[h]
\caption{Transmission Spectra.\label{tbl:tspec}}
\begin{tabular}{cccc}
\tableline\tableline
Bin $\#$ & TrES-2b  &  TrES-4b &  CoRoT-1b\\
& ($R_P/R_\star$) & ($R_P/R_\star$) &($R_P/R_\star$)\\
\tableline
Bin 0 & $0.12850^{+0.00079}_{-0.00079}$ & $0.0932^{+0.0024}_{-0.0024}$ &  $0.1380^{+0.0024}_{-0.0024}$ \\
Bin 1 & $0.12947^{+0.00077}_{-0.00079}$ & $0.0950^{+0.0022}_{-0.0023}$ &  $0.1410^{+0.0029}_{-0.0027}$ \\
Bin 2 & $0.12763^{+0.00096}_{-0.00096}$ & $0.0915^{+0.0022}_{-0.0024}$ &  $0.1384^{+0.0019}_{-0.0020}$ \\
Bin 3 & $0.12881^{+0.00091}_{-0.00091}$ & $0.0939^{+0.0030}_{-0.0030}$ & $0.1410^{+0.0023}_{-0.0023}$ \\
Bin 4 & $0.12997^{+0.00097}_{-0.00097}$ & $0.0967^{+0.0018}_{-0.0018}$ & $0.1389^{+0.0030}_{-0.0045}$ \\
Bin 5 & $0.12855^{+0.00099}_{-0.00099}$ & $0.0943^{+0.0021}_{-0.0021}$ & $0.1410^{+0.0024}_{-0.0024}$ \\
Bin 6 & $0.13000^{+0.00103}_{-0.00103}$ & $0.0942^{+0.0022}_{-0.0022}$ & $0.1396^{+0.0032}_{-0.0040}$ \\
Bin 7 & $0.12955^{+0.00104}_{-0.00104}$ & $0.0954^{+0.0019}_{-0.0019}$ &  $0.1370^{+0.0029}_{-0.0036}$ \\
Bin 8 & $0.12885^{+0.00094}_{-0.00094}$ & $0.0915^{+0.0024}_{-0.0026}$ & $0.1307^{+0.0023}_{-0.0024}$ \\
Bin 9 & $0.12847^{+0.00113}_{-0.00114}$ & $0.1008^{+0.0027}_{-0.0027}$ & $0.1319^{+0.0020}_{-0.0020}$ \\
\tableline
\end{tabular}
\tablecomments{The wavelengths corresponding to each wavelength bin vary by object, and can be found in Table~\ref{tbl:binboundaries}.}
\end{table}

We detect no clear sources of opacity in the atmospheres of TrES-2b, TrES-4b, and CoRoT-1b. Comparing the precision of our observations to $\delta$, we find that we can rule out at $3 \sigma$ spectral variations of 6.3 scale heights for TrES-2b, 8.4 scale heights for TrES-4b, and 9.1 scale heights for CoRoT-1b.  \citet{Tinetti2010} derived a model for the atmosphere of XO-1b featuring variations in absorption of $1.65-1.77\%$ for $\delta=0.00012$, corresponding to variations of 10 scale heights. Based on our sample, such planetary atmospheres are not common. We also note that the recent WFC3 spectrum of XO-1b by \citet{Deming2013} does not confirm the large variations reported by \citet{Tinetti2010}. 

The spectra of TrES-2b and TrES-4b can  be reproduced by a wide range of cloud-free atmospheric models. Our non-detection of a wavelength-dependent source of opacity means we cannot differentiate between these models. These data are also well-fit by a flat line, implying there are also a wide variety of models with haze or clouds that could potentially explain these spectra. We did not explore this hypothesis because the wavelength domain is relatively small, and that measurements over large bandpasses would be necessary to firmly confirm this. The spectrum of CoRoT-1b is not well fit by either carbon-rich or solar-composition atmospheric models, nor a flat line. A natural step forward would be to re-observe this target with a complete baseline (usable pre- and post-transit orbits) to enable use of the higher-precision $divide-oot$ methodology: this would additionally enable further testing of the $single-oot$ methodology used to derive this spectrum.

\subsection{WASP-4b Thermal Emission}

\begin{figure}[h]
\centering
\includegraphics[width=16 cm, angle=0]{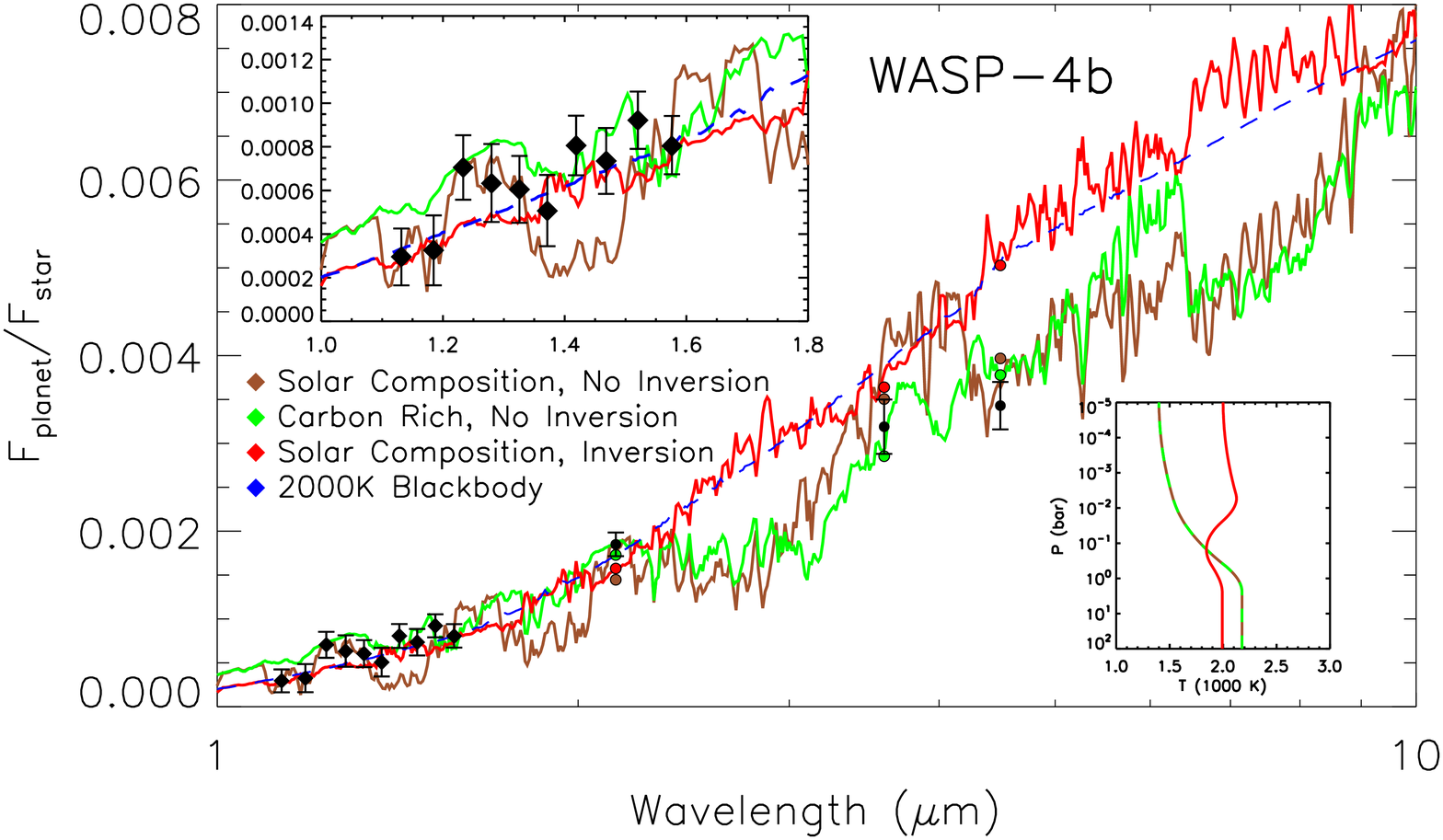}
\caption{WASP-4b thermal emission spectrum. Also plotted are results from past broadband eclipse measurements in \citet{Beerer2011} and \citet{Caceres2011}. Overplotted are a solar-composition atmosphere (brown) and a carbon-rich atmosphere (green) that lack thermal inversions, and a thermally inverted solar-composition atmosphere (red). A 2000 K blackbody is plotted in blue. The colored points show the bandpass-integrated models. The top left inset shows the WFC3 data derived in this paper. The bottom right inset shows the temperature/pressure profiles associated with the model atmospheres: the non-thermally-inverted atmospheres in brown and green, and the thermally inverted atmosphere in red. \label{fig:wasp4emission-10bincut}}
\end{figure} 

Figure~\ref{fig:wasp4emission-10bincut} presents the thermal emission spectrum for WASP-4b in the context of past broadband measurements and representative cloud-free models.  Table~\ref{tbl:espec} presents the spectra in tabular form. Our WFC3 data are consistent with isothermal atmospheres, and atmospheres with and without a temperature inversion. Combining our data with published Spitzer broadband measurements gives us leverage on the thermal state of the atmosphere. An isothermal atmosphere (i.e. a blackbody) fits the WFC3 data with $\chi^{2}=7.92$ with 10 degrees of freedom, and a marginally-inverted solar-composition atmosphere fits the WFC3 data with $\chi^{2}=12.2$, but both are inconsistent with the longer wavelength Spitzer observation. We do not detect excess absorption or emission in the band corresponding to the 1.4$\mu$m water feature. Hence it is unsurprising that a non-inverted solar-composition atmosphere, which would be expected to exhibit water in absorption, fits the WFC3 data comparatively poorly, with $\chi^{2}=31.3$.  A non-inverted carbon rich atmosphere fits the WFC3 data with $\chi^{2}=10.6 $ and is also consistent with broadband measurements. Overall, in conjunction with the Spitzer results the data favor non-inverted water-poor atmospheres for WASP-4b. This result is consistent with the finding of \citet{Beerer2011} that WASP-4b does not have a strong thermal inversion.

\subsection{TrES-3b Thermal Emission}

\begin{figure}[h]
\centering
\includegraphics[width=16 cm, angle=0]{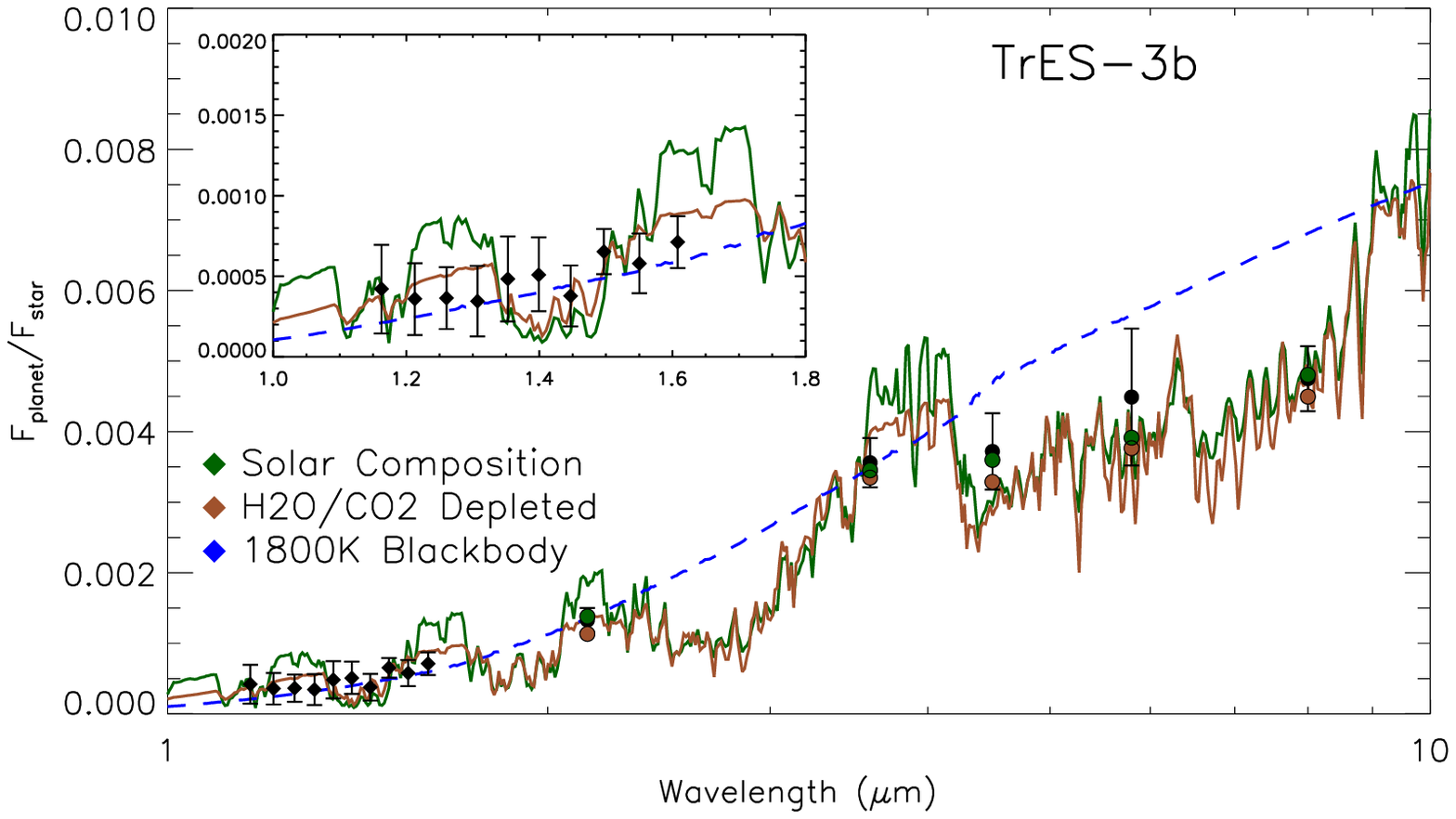}
\caption{TrES-3b thermal emission spectrum. Also plotted are broadband measurements of the eclipse depth in the IR from \citet{Croll2010} and \cite{Fressin2010}. In green is a solar-composition atmospheric model. In brown is a low-metallicity atmospheric model depleted in CO$_2$ and H$_2$O by a factor of 10 relative to solar. Plotted in blue is a blackbody, which is strongly inconsistent with the data. \label{fig:tres3specmodel}}
\end{figure}

Figure~\ref{fig:tres3specmodel} presents the thermal emission spectrum for TrES-3b in the context of representative atmospheric models and past photometric measurements.  Table~\ref{tbl:espec} presents the spectra in tabular form. A solar-composition model fits the WFC3 data with $\chi^{2}=30.4$ with 10 degrees of freedom, indicating it is a poor explanation of the data. This is unsurprising as we do not detect absorption or emission corresponding to the 1.4 $\mu$m water feature in this spectrum. An 1800-K blackbody fits our data very well with $\chi^{2}=3.50$, suggesting an isothermal atmosphere. However, the blackbody is inconsistent with existing Spitzer photometry that requires the presence of absorbing molecules. An atmospheric model depleted in CO$_2$ and H$_2$O by a factor of 10 relative to solar fits the WFC3 data with $\chi^{2}=7.51$. This model is also consistent with broadband photometry. Overall, in conjunction with the Spitzer results the data favor water-poor atmospheres for TrES-3b.

\begin{table}[h]
\small
\caption{Dayside-Integrated Emission Spectra.\label{tbl:espec}}
\begin{tabular}{ccc}
\tableline\tableline
Bin $\#$ & TrES-3b  &  WASP-4b\\
& ($F_{P}/F_{\star}\times10^4$) & ($F_{P}/F_{\star}\times10^4$) \\
\tableline
Bin 0 & $4.2^{+2.7}_{-2.8}$ & $3.0^{+1.3}_{-1.3}$\\
Bin 1 & $3.6^{+2.2}_{-2.2}$ & $3.2^{+1.6}_{-1.6}$\\
Bin 2 & $3.6^{+1.9}_{-1.9}$ & $7.1^{+1.5}_{-1.5}$\\
Bin 3 & $3.5^{+2.2}_{-2.2}$ & $6.3^{+1.8}_{-1.8}$\\
Bin 4 & $4.8^{+2.6}_{-2.6}$ & $6.0^{+1.5}_{-1.5}$\\
Bin 5 & $5.1^{+2.3}_{-2.3}$ & $5.1^{+1.6}_{-1.6}$\\
Bin 6 & $3.8^{+1.9}_{-1.9}$ & $8.1^{+1.4}_{-1.4}$\\
Bin 7 & $6.5^{+1.4}_{-1.4}$ & $7.4^{+1.5}_{-1.5}$\\
Bin 8 & $5.8^{+1.9}_{-1.9}$ & $9.2^{+1.3}_{-1.3}$\\
Bin 9 & $7.1^{+1.6}_{-1.6}$ & $8.0^{+1.4}_{-1.3}$\\
\tableline
\end{tabular}
\tablecomments{The wavelengths corresponding to each wavelength bin vary by object, and can be found in Table~\ref{tbl:binboundaries}.}
\end{table}

\section{Conclusions\label{sec:conc}}
We have derived transmission spectra for 3 hot Jupiters and emission spectra for 2 using the WFC3 instrument on HST. In cases where out-of-transit orbits are available before and after the event, we are able to decorrelate the data entirely without reference to an instrument mode ($divide-oot$). In cases where only one out-of-transit orbit is available, we have validated a minimal two-parameter linear decorrelation to whiten the data, at the cost of significantly reducing parameter estimate precision ($single-oot$). We have demonstrated these decorrelation techniques to be consistent with each other, as well as with an entirely different instrument parameter-dependent decorrelation technique ($differential-eclipse$). We derive consistent spectra with and without flat-fielding and with different choices for background estimation method.

TrES-2b, TrES-4b, and CoRoT-1b are featureless to the precision of our data. Transmission spectra of TrES-2b and TrES-4b are well fit by models of both solar-composition and carbon-rich atmospheres. However, our precision is not high enough to differentiate between these cases. CoRoT-1b is not well fit by either model, and as such particularly merits follow-up observations with an enhanced out-of-transit baseline to enable use of the higher-precision $divide-oot$ methodology. Our WASP-4b transmission spectrum is nonrobust: different treatments of the saturation give significantly different results. Hence we do not report results for this dataset. Follow-up observations of WASP-4b, perhaps using an alternative observation strategy like spatial scan mode, or exposing to levels well below nonlinearity, are required to derive this planet's transmission spectrum.

Our emission spectra of TrES-3b and WASP-4b do not show evidence of water, implying either isothermal atmospheres or atmospheres depleted in water. Taken in context with previous broadband measurements of the eclipse depth, isothermal atmospheres are disfavored. A carbon-rich atmosphere is consistent with the WASP-4b emission spectrum, while a low-metallicity atmosphere is consistent with the TrES-3b emission spectrum.

Our $1-\sigma$ precision in each of the ten bandpasses corresponds to variations of 2.1, 2.8, and 3.0 scale heights for TrES-2b, TrES-4b, and CoRoT-1b respectively. We can rule out atmospheric variation at the level of 10 scale heights and above at $3\sigma$ for all of our planets. Based on this sample, atmospheric models of the kind reported for XO-1b by \citet{Tinetti2010} are not common. Increased precision is required to resolve water on the hot Jupiters we have studied here. Multi-visit observing campaigns obtaining multiple eclipses and transits of a single object are an obvious way to achieve increased precision, especially given that our observations are dominated by photon noise. Our analysis of WASP-4b in transit demonstrates that saturated data may obfuscate analysis; future programs may wish to consider exposing their data to levels well short of nonlinearity to sidestep challenges of reduction. Our work suggests future campaigns should be planned to include orbits before and after the event, as done in other works (e.g. \citealt{Berta2011}, \citealt{Huitson2013}, \citealt{Swain2013}).  Photon collection efficiency can also be improved via the spatial scan mode, which optimizes duty cycle by continuously imaging. \citet{McCullough2012} presents recommendations for observing programs using spatial scan mode, and programs like those of \citet{Deming2013}, \citet{Knutson2014}, and \citet{Kreidberg2014} have been able to use these techniques to extract high-precision atmospheric spectra. For example, \citet{Deming2013} are able to achieve precisions of 35 ppm for HD209458b using this technique. Such campaigns by WFC3 promise to open a new age in the characterization of exoplanet atmospheres.

\acknowledgements
Based on observations made with the NASA/ESA Hubble Space Telescope, obtained at the Space Telescope Science Institute (STCSci). These observations are associated with program HST-GO-12181. Support for this program was provided by NASA through a grant from the Space Telescope Science Institute, which is operated by the Association of Universities for Research in Astronomy, Inc., under NASA contract NAS 5-26555. This material is based upon work supported by the National Science Foundation Graduate Research Fellowship under Grant No. DGE-1144152. This work was also performed in part under contract with the California Institute of Technology (Caltech) funded by NASA through the Sagan Fellowship Program grant awarded to JMD. NM acknowledges support from Yale University through the YCAA postdoctoral prize fellowship. This research has made use of the Exoplanet Orbit Database and the Exoplanet Data Explorer at exoplanets.org; the SIMBAD database, operated at CDS, Strasbourg, France; NASA's Astrophysics Data System Bibliographic Services; and SAOImage DS9, developed  by Smithsonian Astrophysical Observatory. The authors are grateful to Z. Berta and J. Carter for many fruitful discussions, to the STSci HST help team for their assistance, and to an anonymous referee whose comments strengthened the paper.

\emph{Facilities}: HST (WFC3)
\bibliography{msv3b}{}

\end{document}